\documentclass[aps,prc,twocolumn,preprintnumbers,showpacs,amsmath,amssymb,floatfix]{revtex4}

\usepackage{graphicx}
\usepackage{dcolumn}
\usepackage{bm}
\usepackage{epsfig}
\usepackage{xspace}
\usepackage{rotating} 

\begin{document}
\newcommand{\singlewidth}{6.8in}
\newcommand{\doublewidth}{3 in}
\newcommand{\figwidth}{\doublewidth}

\newcommand{\degree}{\ensuremath{^{\circ}}\xspace}
\newcommand{\degrees}{\degree}
\newcommand{\dsdo}{\ensuremath{{ \frac{d\sigma}{d\Omega}}} }
\newcommand{\ddp}{\ensuremath{\rm{^2H}({\it d},{\it p}){^3H}}\xspace}
\newcommand{\sddp}{\ensuremath{\rm{^2H}({\it d},{\it p})}\xspace} 
\newcommand{\ddn}{\ensuremath{\rm{^2H({\it d},{\it n})}{\rm^3He}}\xspace} 
\newcommand{\sddn}{\ensuremath{\rm{^2H({\it d},{\it n})}\xspace}} 
\newcommand{\cdd}{\ensuremath{ {}^{12}{\rm C({\it d},{\it d})}}\xspace}
\newcommand{\audd}{\ensuremath{\gold({\it d},{\it d})}\xspace}
\newcommand{\aupp}{\ensuremath{\rm {}^{197}Au({\it p},{\it p})}\xspace}
\newcommand{\saudd}{\ensuremath{\rm Au({\it d},{\it d})}\xspace}
\newcommand{\saupp}{\ensuremath{\rm Au({\it p},{\it p})}\xspace}
\newcommand{\ddd}{\ensuremath{ \rm{^2H}({\it d},{\it d})}\xspace}
\newcommand{\pdd}{\ensuremath{\rm {^1H}({\it d},{\it d})}\xspace}
\newcommand{\pdp}{\ensuremath{\rm {^1H}({\it d},{\it p})}\xspace}
\newcommand{\ppp}{\ensuremath{\rm {^1H}({\it p},{\it p})}\xspace}
\newcommand{\cpp}{\ensuremath{\rm {}^{12}{C}({\it p},{\it p})}\xspace}
\newcommand{\cdp}{\ensuremath{\rm {}^{12}{\rm C}({\it d},{\it p})}\xspace}
\newcommand{\dpp}{\ensuremath{ \rm{^2H}({\it p},{\it p})}\xspace}
\newcommand{\sddhet}{\ensuremath{ \rm{^2H}({\it d},{}^3{\rm{He}})}\xspace}
\newcommand{\ddhet}{\ensuremath{ \rm{^2H}({\it d},{}^3{\rm{He}}){\it n}}\xspace}
\newcommand{\sdpp}{\ensuremath{ \rm{^2H}({\it p},{\it p})}\xspace}
\newcommand{\dpd}{\ensuremath{\rm{^2H}({\it p},{\it d})}\xspace}
\newcommand{\ddt}{\ensuremath{{\rm ^2H}({\it d},{\it t}){\rm ^1H}}\xspace}
\newcommand{\dsig}{\ensuremath{\sigma(\theta)}\xspace}
\newcommand{\hethree}{\mbox{$^{3}$He}\xspace}
\newcommand{\hethrees}{\mbox{$^{3}$He's}\xspace}
\newcommand{\hthree}{\mbox{$^{3}$H}\xspace}
\newcommand{\hefour}{\mbox{$^{4}$He}\xspace}
\newcommand{\dst}[2]{\ensuremath{\sigma_{#1}{#2}(\theta)}\xspace}
\newcommand{\dste}[2]{\ensuremath{\sigma_{#1}{#2}(E,\theta)}\xspace}
\newcommand{\dsteb}[3]{\ensuremath{\sigma_{#1}{#2}(E_{#3},\theta)}\xspace}
\newcommand{\gold}{\ensuremath{^{197}{\rm Au}}\xspace}
\newcommand{\carb}{\ensuremath{^{12}{\rm C}}\xspace}
\newcommand{\eqr}[1]{\begin{equation}\begin{array}{lcl}#1\end{array}\end{equation}}
\newcommand{\enum}{\parbox[b]{1cm}{\begin{eqnarray}\end{eqnarray}}}
\newcommand{\alp}{\ensuremath{\alpha^p}\xspace}
\newcommand{\ad}{\ensuremath{\alpha^d}\xspace}
\newcommand{\ar}{\ensuremath{\alpha^r}\xspace}
\newcommand{\ax}{\ensuremath{\alpha^x}\xspace}
\newcommand{\rate}{\ensuremath{N_a\left<\sigma\nu\right>}\xspace}
\newcommand{\dpg}{\ensuremath{\rm{^2H}({\it p},\gamma)}\xspace}

\title{Precision Measurements of $\rm{^2H}({\it d},{\it p})^3H$ and $\rm{^2H}({\it d},{\it n})^3{\rm He}$ Total Cross Sections at Big-Bang
Nucleosynthesis Energies}

\author{D.~S. Leonard}
\altaffiliation{Present address: University of Alabama, Tuscaloosa, AL}
\author{ H.~J. Karwowski}
\author { C.~R. Brune}
\altaffiliation{Present address: Ohio University, Athens, OH}
\author{ B.~M. Fisher}
\altaffiliation{Present address: Tulane University, New Orleans, LA}
\author{ E.~J. Ludwig}
\affiliation{Department of Physics and Astronomy, 
        University of North Carolina at Chapel Hill, U.S.A. \\ and  
        Triangle Universities Nuclear Laboratory, 
        Durham, North Carolina}

\date{\today}

\begin{abstract}
Recent  Wilkinson Microwave Anisotropy Probe (WMAP) measurements have determined the baryon density of the Universe $\Omega_b$ with a precision of about 4\%\@. With $\Omega_b$ tightly constrained, comparisons of Big Bang Nucleosynthesis (BBN) abundance predictions to primordial abundance observations can be 
made and used to test BBN models and/or to further constrain abundances of isotopes with weak observational limits. To push the limits and 
improve constraints on BBN models, uncertainties in key nuclear reaction rates must be minimized. To this end, we made new precise 
measurements of the \ddp and \ddn total cross sections at lab energies from 110~keV to 650~keV\@.  
  A complete fit was 
performed in energy and angle to both angular distribution and normalization data for both reactions simultaneously.  By including  
parameters for experimental variables in the fit,  error correlations between detectors, reactions, and reaction energies were accurately tabulated by 
computational methods.  With uncertainties around $\rm 2\% \pm 1\%$ scale error,  these new measurements significantly improve on the existing data set.  At relevant temperatures, using the data of the present work, both reaction rates  are found to be about 7\% higher than those in the widely used Nuclear Astrophysics Compilation of Reaction Rates (NACRE).  These data will thus lead not only to reduced uncertainties, but also to modifications in the BBN abundance predictions. 

\end{abstract}
\maketitle

\section{Introduction}
\label{chap:intro}
 
The standard Big Bang model explains remarkably well many
features of the Universe which are otherwise difficult to reconcile.  
The standard Big-Bang nucleosynthesis (BBN) model consists of a small network of nuclear reactions occuring at energies easily obtained in the lab. The outcome of Standard BBN is
determined almost entirely by the nuclear reaction rates and the baryon to
photon ratio, $\eta$, of the Universe, which is directly related to the baryon density $\Omega_b$ ($\Omega_bh^2=3.66\times10^7 \eta$, where $h$ is the Hubble constant in units of $\rm 100 km\times s^{-1}\times Mpc^{-1}$~\cite{PDG} ).  With detailed network calculations and knowledge of the nuclear 
cross-sections, all of the primordial abundances can be precisely
calculated as a function of $\eta$. 
Knowledge of any one abundance or a separate determination of $\eta$ thus allows all other primordial abundances to be inferred from the Standard BBN model.  Knowledge of any two of these observables produces a check of the model itself.  

In the past, quantitative understanding of BBN has been limited by uncertainties in the observed
primordial abundances and the value of $\eta$.  Until recently $\eta$ was treated as a free parameter.  As observations have improved, the value of $\eta$ has become a well determined input and is instead used, along with the nuclear reaction rates, to predict the primordial abundances.  Regardless of which values are assumed and which are predicted, the uncertainties of the cross sections themselves are becoming a significant factor in precision tests of BBN models.  Improved analyses depend on having accurate and precise knowledge of the nuclear reaction rates, their uncertainties, and the correlations in those uncertainties.
 
\subsection{Primordial Deuterium Observations} Recent measurements of
absorption lines in high red-shift, metal poor, QSO-back-lit gas clouds
have constrained the primordial deuterium abundance $D$ to the impressive
interval of ${D/H}=2.78_{-.38}^{+.44}\times10^{-5}$~\cite{kirkman}, expressed relative to the hydrogen abundance $H$\@.  The measurement and analysis procedure is well described in the review article of Tytler {\it et al.}~\cite{Tyt00}\@. 
Currently systematic scatter limits the precision of the deuterium abundance observations~\cite{kirkman}, but as more data arrive and the systematics become better understood this could quickly change.  

Measurements of primordial deuterium abundances \cite{bur98a,bur98b}, along with recent measurements of $\eta$ from WMAP data~\cite{cyburt,spergelwmap}, bring the nuclear reaction rates increasingly closer to being the limiting factors in testing the consistency of the Standard BBN model.

\subsection{Cosmology Enters the Lab} 

The value of $\eta$ determined from the WMAP results is in good agreement with primordial deuterium abundance measurements of Kirkman~{\it et al.}~\cite{kirkman}\@.  However limited amounts of data and the systematic uncertainties in abundance measurements and in reaction rates have limited the level of precision at which the BBN models can be tested to around 10\% for most observables.

\begin{figure}[htb]\centering
\includegraphics[width=\figwidth]{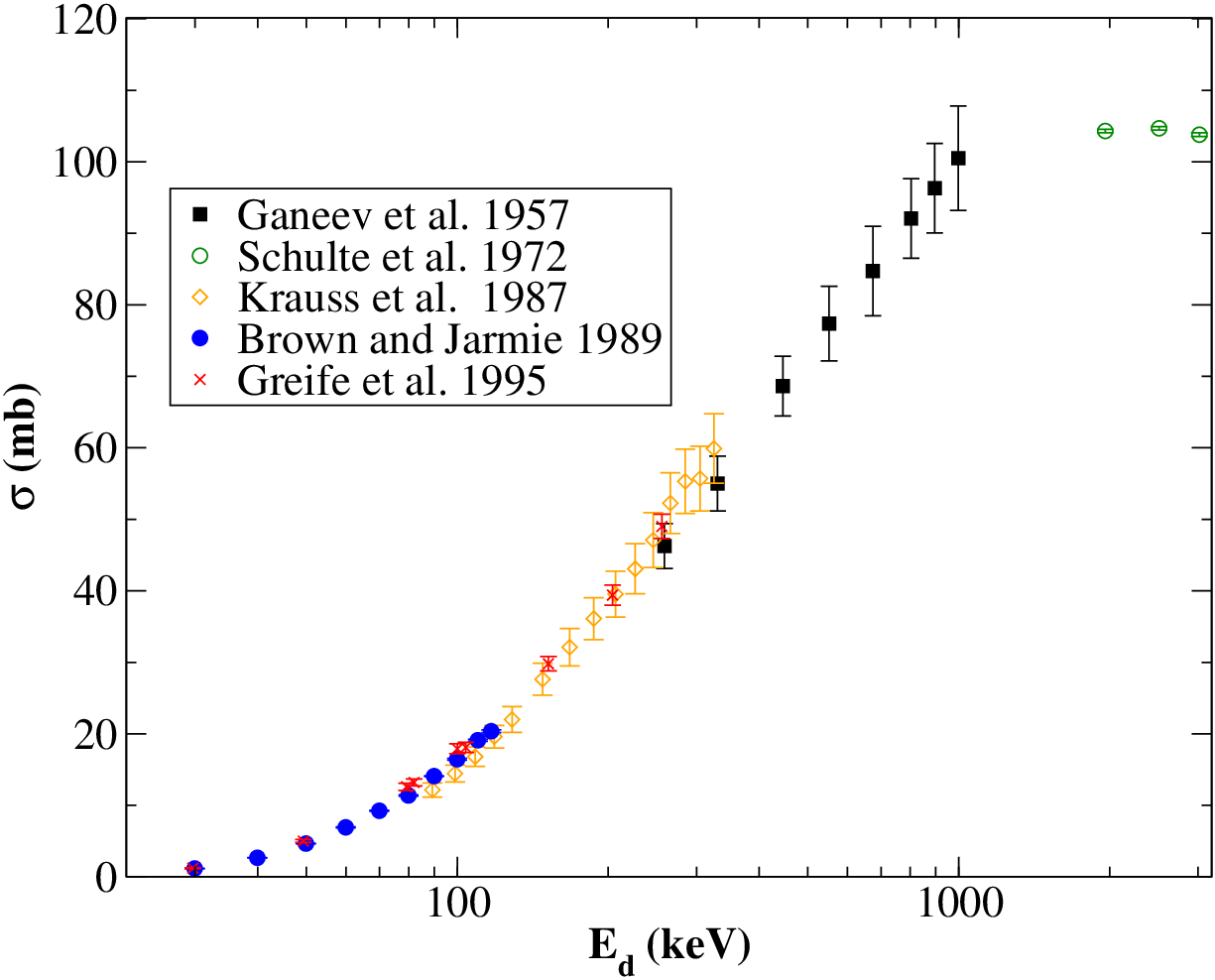}
\caption[Prominent data sets for \ddn cross-sections at BBN energies.]{(Color) Prominant data sets for \ddn cross-sections at BBN energies}
\label{fig:ddndata}
\end{figure}

With the uncertainties of previously existing data, the \ddp and
\ddn reaction cross-sections at energies in the range of a
few hundred keV make large contributions to the uncertainties in deuterium
abundances as predicted by network calculations~\cite{nollettburles,bur99a,bur99b,sch98}\@.  Nollett and Burles~\cite{nollettburles} provide sensitivity functions estimating contributions to the deuterium and $\rm ^7Li$ abundances of several reactions as a function of energy.  The most relevant energy range of both the \ddn and \ddp reactions extends from roughly $\rm {\it E_d}=100~keV$ up to about $\rm {\it E_d}=700~keV$.  Prior to our measurements, high precision data for these reactions were very limited in this range, as shown in Fig.~\ref{fig:ddndata}\@.    The \ddp reaction data is qualitatively very similar.   From $\rm {\it E_d}=325~keV$ to the top of the BBN energy range there were very few data points, all having large uncertainties.  Even at lower energies of significance to BBN, the uncertainties of previously existing data were around the 10\% level or only slightly better.   For the energy range above 325~keV, one of the more complete data sets was that of Ganeev {\it et al.}~\cite{ganeev}\@.   These data are not included in the NACRE data compilation~\cite{NACRE},  considered the most prominent collection of experimental rates for reactions of astrophysical significance.

In order to aid in tightening the BBN constraints, we measured total cross sections for both of these deuterium burning reactions at lab energies ranging from about 112~keV to 646~keV\@.  By carefully establishing experimental procedures in order to maximally cancel the dependencies of yields on experimental parameters, we have obtained about 2\% statistical uncertainties plus or minus a 1\% systematic scale uncertainty.  Furthermore our $\chi^2$ analysis techniques have allowed optimal use of the data and  provided correlations in the uncertainties of the two reactions across the range of energies.  The new data form a significant improvement of the inputs to the BBN calculations and facilitate an emerging era of high-precision BBN.  

\section{Experiments}
\label{chap:exp}

To determine the integrated cross sections for the \ddp and \ddn reactions, relative angular distributions of the differential cross sections \dsdo were measured, henceforth denoted as simply \dst{}{}, or \dste{}{}, where $E$ and $\theta$ represent the energy and reaction angle respectively.  We normalized these distributions by measuring absolute differential cross sections at selected fixed angles.  This general procedure was performed for eight energies, ${E_d}$=120, 180, 240, 320, 390,  480, 560, and 650~keV\@.  These are approximate nominal energies and are the values used here to refer to the various data sets.  The determination of the exact energies will be discussed in Sec.~\ref{sec:exp:energetics}\@. Generally data for both reactions were obtained simultaneously.  

All data were taken using the TUNL Low-Energy Beam Facility (LEBF) and the High-Voltage Target Chamber~\cite{hv}\@.  The LEBF is composed of the Atomic Beam Polarized Ion Source~\cite{cle95} and Mini-Tandem accelerator~\cite{mt}\@.  The combined acceleration potentials of the source, Mini-Tandem and HV chamber provide deuteron and proton beam energies up to approximately $\rm 680$~keV\@. All targets used were carbon-based, self-supporting transmission targets. 

\subsection{Normalization Technique}
\label{exp:norm}
\label{exp:norm:solution}
\label{exp:norm:procedure}

For the present measurements, the central technique used to determine the absolute cross sections was to compare them to those of a reference reaction.  
We used $p$--$d$ elastic scattering for which
multiple absolute measurements of differential cross sections exist at energies near the range of interest for our measurements~\cite{woo02,brunepd,huttelpd}\@.  The data of Brune {\it et al.}~\cite{brunepd} was taken with the explicit purpose of normalizing relative differential cross-section data taken with the same equipment as the present work.  All of these measurements agree, to within about 1\%, with theoretical few-body calculations with no free parameters~\cite{kie95,woo02}\@.  
We used the calculated values as the cross-section reference for our measurements.
 
In order to compare yields for the $d$--$d$ interactions and $p$--$d$ interactions under similar conditions, we alternated between deuteron beam and proton beam incident on the same self-supporting, deuterated carbon targets.  The targets were produced by plasma deposition of fully-deuterated methane gas and were typically about 30~$\mu$g/cm$^2$ thick, with roughly equal numbers of carbon and deuterium atoms.

The ratio of cross sections for the two observed interactions is given by
\begin{equation}
\frac{\dsteb{d}{}{d}}{\dsteb{p}{}{p}}
  = \frac{n_pt_p \Delta\Omega_p }
   {n_dt_d \Delta\Omega_d }\frac{N_d}{N_p}\,,
\label{eqn:sigratio}
\end{equation}
where $n$ is the incident number of beam particles, $N$ is the detected number of particles, and $t$ is the areal density of target nuclei.  The subscripts $p$ and $d$ refer to measurements with proton and deuteron beam respectively. By using the same experimental setup and target for both beams, the solid angles and target thicknesses were made to cancel.  

One of the larger obstacles in a direct cross-section measurement is measuring the time-integrated beam flux incident on the target.  
For the energies of the present experiment,  the determination of the number of incident beam particles by charge collection and integration is made difficult by angular straggling and charge exchange in the target.

In order to determine the ratio of proton beam particles to deuteron beam particles, a thin layer of gold, approximately 10 or 20~\AA~thick (1\AA~of gold is equivalent to $\rm 1.7\times10^{15}~atoms/cm^2$), was evaporated onto one surface of the target.  The gold layer was then oriented on the upstream side of the target, facing the incident beam.  A back-angle detector was used to monitor Rutherford elastic back-scattering from gold for both proton and deuteron beams.  

The ratio of the number of incident beam particles for the two beams is given by 
\\\begin{equation}
\frac{n^p}{n^d}=\frac{N^p}{N^d} \frac{\sigma_{\saudd}(E_d,\theta)}{\sigma_{\saupp}(E_p,\theta)}\,,
\end{equation} 
where $\sigma_{\saudd}(E_d,\theta)$ and $\sigma_{\saupp}(E_p,\theta)$ are respectively the differential cross sections for proton and deuteron scattering from gold at the same lab angle. 

When beam is incident on the amorphous deuterium targets, deuterium is depleted from them, albeit slowly.  If a target is depleted in a spatially non-uniform manner then two beams striking slightly different areas at different times may not interact with a target of the same thickness.    
High intensity beam currents tend also to produce macroscopic defects in targets such as rips and holes, especially at these energies where energy loss in the target is large.  Depletion and stability problems seem to depend not only on the total number of particles which pass through the target, but also on the rate at which they pass. 

A constant or slowly varying target thickness must be maintained between proton and deuteron runs. Since the reactions of interest have cross sections on the order of $\rm mb/sr $, as opposed to $\rm b/sr$ for elastic scattering, a much larger integrated beam flux was required in order to acquire sufficient reaction data.  Such high beam fluxes would produce prohibitively large target variations.       

To avoid this problem, we first normalized $d$--$d$ elastic-scattering yields to the $p$--$d$ yields using small beam currents, below 10~nA.  At these currents target depletion was almost imperceptible,  no more than 1\% per hour, and many targets were structurally stable for several hours.  Data was taken later with higher beam currents, around 50 to 100~nA, and the ratio of $d$--$d$ elastic-scattering yields to \ddp and to \ddn reaction yields was obtained.  Since the elastically scattered deuterons and the reaction products were measured simultaneously in this last step, they passed through the same target thicknesses.  This final step adds negligibly to the overall systematic error of the measurement.  To further ameliorate the depletion problem, the beam switching technique was employed several times over short intervals, and small entrance collimators were used to define precisely the incident beam trajectory.  

In order to change beam types rapidly, we injected a mixture of deuterium and hydrogen gas into the ECR ionizer of the Polarized Ion Source~\cite{cle95}\@.  
With this dual-beam source configuration, by changing only the inflection-magnet current, we could easily put over 100 nA of either beam on target. 

\subsection{Experimental Setups and Procedures}
\label{sec:exp:chamber}
\begin{figure}[!ptbh] \centering
    \includegraphics[width=\figwidth]{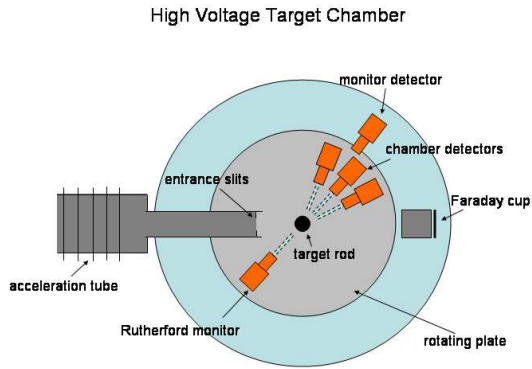}
    \caption[The HV Scattering Chamber.]{(Color online) Schematic of the experimental set-up. The entire scattering-chamber can be raised to a potential of $\pm$200~kV to accelerate or decelerate the beam.}
    \label{fig:exp:hv}
\end{figure}

The HV chamber which was used for these measurements is described in Ref.~\cite{hv}\@.  The chamber's basic features are illustrated in Fig.~\ref{fig:exp:hv}\@.  
Two independently rotating plates are installed, one on the top of the chamber and one on the bottom, which can be used to mount detectors on the left and right side of the chamber respectively.   The fixed monitor detectors are placed above and below the reaction plane having a view of the target which was not obstructed by the rotating detectors.  The beam enters the chamber through an acceleration tube allowing for an acceleration through a 200 kV potential.  It then passes through vertical and horizontal entrance slits which define the beam position.  

\subsubsection{High-Energy Normalization Setup}
\label{exp:norm:setup}
\begin{figure}[pb!] \centering
\vspace{.1in}
    \includegraphics[width=\figwidth]{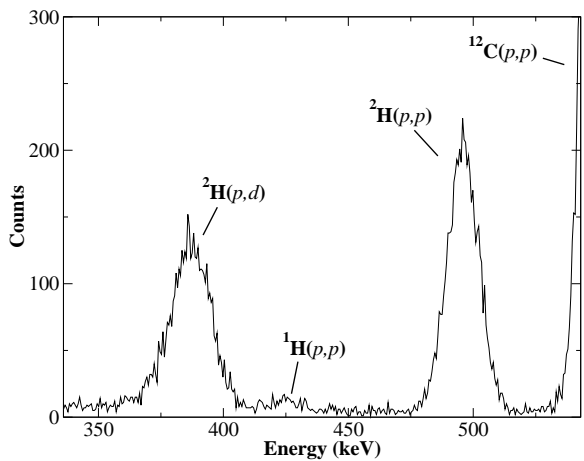}
   \caption[Spectrum of proton elastically scattering on a deuterated target.]{Spectrum from proton elastic scatterring from a deuterated target at $\theta_{\rm lab}=27\degrees$, $E_p=$560~keV\@.}
    \label{fig:exp:pdspec}
\end{figure}
\begin{figure}[pbth] \centering
\vspace{.1in}
    \includegraphics[width=\figwidth]{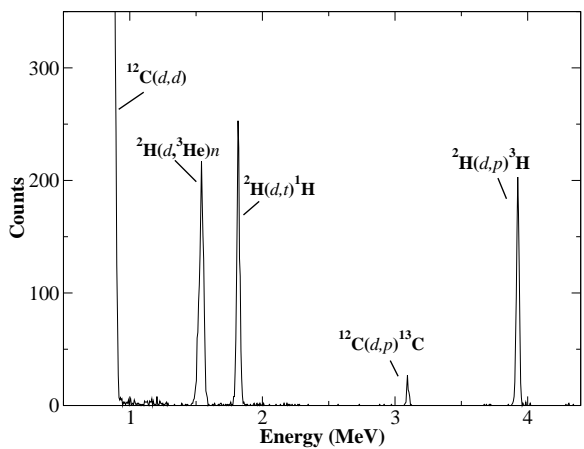}
    \caption[Spectrum of \it{d}+\it{d} reaction products.]{Spectrum of \it{d}+\it{d} reaction products at $\theta_{\rm lab}=35\degrees$, $\rm{\it  E_d}=480~keV$. }
    \label{fig:exp:reacspec}
\end{figure}

For the highest four beam energies, 390, 480, 560, and 650~keV, we used the procedure described in Sec.~\ref{exp:norm:solution} to normalize the differential cross sections.    Seven high-resolution, ion-implanted silicon detectors were positioned in the chamber to observe elastic scattering from deuterium.  All detectors had either 300 or 500 $\mu$m depletion depths, and 1.50~$\rm cm^2$ active areas.  Four were placed on the rotating detector tracks and set at 35.0\degrees and 48.0\degrees on both the left and right sides.  Two more were mounted as out-of-plane monitors at 27.5\degrees.  One detector was placed on the top rotating plate at 125.0\degrees left for the purpose of monitoring Rutherford backscattering from gold as explained in Sec.~\ref{exp:norm:procedure}\@.  The rotating detectors were placed approximately 18~cm from the target with collimators directly in front of them with openings 0.46~cm wide and 0.95~cm tall.  The out-of-plane and back-angle monitor detectors were mounted at approximately 30~cm and 10~cm from the target respectively.   The out-of-plane monitors were given collimators 0.64~cm wide by 0.95~cm  tall, and the back-angle monitor was equipped with a circular collimator of 1.27~cm diameter. Cylindrical aluminum tubes were placed in front of all detector collimators to help reduce possible background scattering from various sources such as the entrance slits. 
Small permanent magnets were placed around the tubes to reduce the effects of electrons released from the target.  Systematic cancellations inherent in the procedure removed the need for precise knowledge of collimator sizes and radii.  An aluminum plate serving as a Faraday cup, was placed at the back of the chamber to integrate the beam current.  Current integration was used for beam adjustment and diagnostic purposes only. 

The two pairs of detectors on the rotating plates were used to observe $p$--$d$ elastic-scattering yields during proton beam runs. A spectrum is shown in Fig.~\ref{fig:exp:pdspec}\@. 
The out-of-plane monitors were used to measure $d$--$d$ elastic-scattering yields during the interleaved deuteron runs.  The in-plane detectors were again used for observing reaction products (see Fig.~\ref{fig:exp:reacspec}) during the reaction runs while simultaneously measuring $d$--$d$ elastic-scattering yields in the out-of-plane monitors.  Using different detectors for the two simultaneous deuterium-beam measurements provided more flexibility for the choice of angles used and more easily facilitated differing signal gain requirements for the elastic-scattering and reaction products.  Solid angle cancellations required that all $d$--$d$ elastic yields were measured in the same dedicated detectors and that the reaction yields were measured in the same detectors as the $p$--$d$ elastic scattering yields.

Measurements of the $d$--$d$ scattering yields were complicated by the presence of hydrogen contamination in the targets.  Within the available increments, the angle of 27.5\degrees was experimentally determined to produce the best spectral separation of the $d$--$d$ scattering peak from  both the  \pdp and \cdd peaks.
At $\rm {\it E}_{\rm lab}= 390~keV$, the lowest energy where the beam-switching normalization was performed, adequate spectral  separation of the $d$--$d$ scattering peak from the \pdp peak could not be achieved.  Instead the yield from the well-separated \pdd scattering peak was used to determine the \pdp yield which could then be subtracted from total yield of the unresolved \ddd and \pdp peaks. 
If the ratio of hydrogen to deuterium were constant in every target, the \pdp yield would introduce no error since it would still provide a consistent, proportional measure of the amount of deuterium in the target.  Typically the \pdp yield was less than 10\% of the \ddd yield.  The ratio of the two varied by about 30\% from target to target, thus producing about a 3\% error.  Only a rough correction was needed to reduce this to a negligible effect.
In order to measure the ratio of the \pdd yield to \pdp yield, monitor spectra were collected at several energies for deuteron scattering from a hydrogenated carbon target.

\subsubsection{Low-energy Normalizations}
\label{exp:norm:bootstrap}

It was not possible to normalize data at the lowest energies, 120, 180, 240, and 320~keV using the method described in Secs.~\ref{exp:norm:procedure} and ~\ref{exp:norm:setup}\@.  At these energies it was not possible to resolve all needed elastic peaks with sufficient precision. We instead normalized reactions at these energies directly to reactions at 480~keV using the differential cross sections already obtained for 480~keV from the previous method.
 
For this normalization, four detectors on each side of the chamber were placed at 60\degrees, 44\degrees, 28\degrees, and 13\degrees.  As before, one detector was placed at 164\degrees on the top rotating plate to monitor backscattering from an upstream gold target layer.  All detectors were placed approximately 11~cm from the target and collimated with 1.27~cm diameter circular collimators.

At these low bombarding energies, amorphous deuterated carbon targets usually became very fragile, and were replaced with deuterated parapolyphenol (DPP, chemical formula: $\rm C_{6}D_{4}$ )  targets~\cite{dpptargets}\@.  We constructed the targets by evaporating an approximately 5~$\mu$g/cm$^2$-thick layer of DPP onto a 5~$\mu$g/cm$^2$ carbon foil.  As in the high-energy normalizations, a roughly 3~$\mu$g/cm$^2$-thick Au layer was evaporated onto the DPP surface.  Again the gold layer faced upstream so that the beam first passed through the gold, then the DPP and then the remaining carbon backing.

The actual normalization procedure was fairly straightforward.  We bombarded the target with deuterium beam at 480~keV and observed ratios of reaction yields to Au backscattering yields.  We then changed the beam energy to the energy of interest while using the same target.  The procedure was repeated several times.  Taking the ratios of the two resulting sets of normalized reaction yields and dividing out the two Rutherford cross sections provided the ratios of differential cross sections at 480~keV to the cross sections at each of the lower energies.

\subsubsection{Angular Distributions}
\label{sec:exp:dsdo}

The normalization procedures described above were collectively the most challenging part of the cross section measurements, but they only determined the differential cross section at a few fixed angles.   To measure the relative angular distribution of differential cross sections, the chamber was set up with six detectors placed on the rotating tracks, three on each side, with 13\degrees separation between each detector.  Two pairs of out-of-plane monitor detectors were placed at approximately 10\degrees and 40\degrees.  Detectors were placed roughly 18~cm from the target with circular collimators of 1.27~cm in diameter. Beam currents of 50 to 100~nA were used.  Amorphous carbon targets were used at the highest energies for their high deuterium content. At lower energies carbon backed DPP targets were used for their superior durability.   Using this arrangement, data  at all angles could be normalized to data from the fixed monitor detectors, thus dividing out all beam current and target thickness information and leaving only a relative angular distribution of \dst{}{}\@.

The elastic-scattering rate from carbon present in the target was much higher than the rate of $d+d$ reactions.   With the high Q-values of the reactions, 4.03~MeV and 3.27~MeV for \ddp and \ddn respectively, it was possible to separate the reaction peaks from the lower-energy elastic-scattering peaks for most reaction angles and beam energies.  
To avoid overloading the data acquisition system, Mylar stopping foils from 1 to 6~$\mu$m-thick, depending on the angle and energy, were placed in front of the detectors.

\subsubsection{Detector Electronics}
\label{sec:exp:elec}
As described in Ref.~\cite{hv}, the High-Voltage Chamber accommodated many silicon surface-barrier detectors in various moveable configurations.  For many parts of the experiment, the chamber was electrically isolated from ground and brought to a high negative 
potential of as much as 200~kV.  This required that all energy signals be sent from the chamber via analog fiber-optic transmitters as described in Ref.~\cite{fiber}\@.    Energy signals from the fiber-optic receivers were then routed into six Northern ADC's.   Pulses of known rate were sent through the same electronics in order to measure the deadtime.

\subsection{Energetics}
\label{sec:exp:energetics}
To achieve the desired uncertainties in the cross sections, reaction and scattering energies must be known very well.  
\label{sec:exp:energetics:beam}
The incident beam energy was determined by the potentials on the ion source, on the Mini Tandem, and on the High-Voltage Chamber.  
The energy calibration of the LEBF system was previously determined by a procedure described elsewhere~\cite{hv}\@.  Small corrections were made for energy losses in the Mini Tandem carbon stripping foil and for the potential on the cesium in the charge exchange canal in the ion source.  The uncertainty in the incident beam energy was less than 1~keV\@.

\label{sec:exp:energetics:eloss}
Energy losses in the targets must also be well known in order to determine precisely the reaction and scattering energies.  
In particular the elastic-scattering cross sections used to determine the normalizations are very sensitive to energy, especially at the lowest energies.  
For these low-energy normalizations a three-layer target was used.  The deuteron beam first passed through a thin layer of gold followed by a layer of DPP and finally passed through a carbon backing.   As explained in previous sections, elastic-scattering yields were measured from the gold and the carbon, and reactions were measured from the deuterium in the DPP layer.  Since energy was lost continuously throughout the thickness of the target, the reaction energy was less than the gold scattering energy which was less than the incident beam energy.  Both interaction energies need to be known.

By rotating the multilayer-targets by 180\degrees and observing shifts in the energies of elastically backscattered beam particles, as well as by monitoring relative scattering yields of various targets, it was possible to measure and continuously monitor target thicknesses and energy losses.   The energy losses, due primarily to the carbon content of the targets, did not change significantly over time. Final values of incident beam energies and total energy losses in the gold and deuterated layers are given in Table~\ref{tab:loweloss}\@.  The high energy normalizations were performed on two-layer targets of gold and a deuterated amorphous carbon.   The energy losses in these targets are shown in Table~\ref{tab:higheloss}\@.  The impact of the energy uncertainty is discussed in Sec.~\ref{sec:anal:error:zero}\@.
 
\begin{table*}[htb!]\centering
\caption[Energy corrections for low energy normalizations.]{Energy corrections for low energy normalizations:  The final interaction energy is calculated using the total energy loss in gold and half of the DPP energy loss.  
The values shown here for 480~keV correspond to the target used  to normalize the 120~keV data to the 480~keV data.}
\begin{tabular}{ccccc}
\hline
Nominal  & Incident Deuteron & Total Loss in & Total Loss in & Central Deuterium-\\
Energy (keV) & Energy (keV) &   Gold (keV)& DPP (keV) & Layer Energy (keV) \\      
\hline\hline
480 &  477.0  & 0.5 & 7.1   &   473.0 \\
320 &  317.3  & 0.2 & 5.1   &   314.6 \\
240 &  237.2  & 0.5 & 7.9   &   232.8 \\
180 &  177.3  & 0.5 & 8.1   &   172.8 \\
120 &  116.8  & 0.4 & 8.5   &   112.2 \\
\hline
\end{tabular}
\label{tab:loweloss}
\end{table*}

\begin{table*}[htb!]\centering
\caption[Energy corrections for high-energy normalizations.]{Energy corrections for high-energy normalizations:  Each row describes deuteron beam energies of the data being normalized and the proton energy used to normalize it.  Proton beams having two different energies were used on the same target to normalize the 560 keV deuteron data. Variations in energy losses between rows come from varying stopping powers and variations in target thicknesses.}    
\begin{tabular}{ccccc}
\hline
Nominal & Incident Deuteron & Central Deuteron & Incident Proton & Central Proton\\  
Deuteron Energy      & Beam Energy & Beam Energy & Beam Energy & Beam Energy \\ 
(keV) & (keV) & (keV) & (keV) & (keV)\\     
\hline\hline
650 &  653.2  & 646.1  & 653.6 & 649.0  \\   
560 &  564.4  & 557.3  & 653.6 & 649.4  \\  
560 &         &        & 564.8 & 560.2  \\
480 &  477.0  & 470.2  & 564.8 & 560.8   \\      
390 &  387.4  & 379.2  & 564.8 & 560.4    \\   
\hline  
\end{tabular}
\label{tab:higheloss}
\end{table*}

\section{Data Analysis}
\label{chap:analysis} 

The goal in analyzing the data was to make full use of all the interconnected data sets constraining various physical and experimental parameters  without double counting any statistics, and to understand the correlations in the uncertainties in the final results.   By using a $\chi^2$ analysis to fit all the data to a model constructed from the relevant parameters, it was possible to maximize the amount of information used while appropriately handling and quantifying correlated information and uncertainties.  

Before performing the global fit of the complete data set, some initial analysis was performed on the data from each storage sequence, referred to as a run.  The energy spectra were first analyzed to determine the yields in all peaks of interest. 
When required, fits were made to Gaussian peaks summed with linear or quadratic backgrounds.  
Data were then adjusted for deadtime corrections, monitor normalizations, and cross-section normalizations as needed. In this way a preliminary conventional analysis could be performed, and parameters which changed with each run, primarily including target layer thicknesses and the number of incident beam particles, could be removed from the data sets.

\subsection{High-Energy Normalizations}
\label{sec:analysis:crunch:highnorm}
As explained in Sec.~\ref{exp:norm:procedure}, the highest-energy differential cross sections were normalized to the known $p$--$d$ elastic-scattering cross sections.  Here we outline the procedure needed to obtain absolute reaction cross section from the $p$--$d$ cross sections.   

First we construct the specific yield equations for all reactions and run types required for the calculation. For simplicity we will assume that each yield was observed in only one detector and one data run.  The detector used to measure $p$--$d$ elastic-scattering yields will be referred to as detector A.  The $d$--$d$ elastic-scattering yield was observed in detector B in a beam switching run.  These two runs used the same target, which we will call target 1, having a deuterium target thickness $t_{d1}$ and a gold target thickness $t_{g1}$. Finally $d$--$d$ elastic scattering was measured in detector B simultaneously with the $d$--$d$ reactions in detector A.  These measurements were made using a different target, target 2. For all three measurements, elastic scattering on gold was measured in detector C\@. The superscript $p$ indicates a proton-beam run, $d$ a deuteron beam-switching run, and finally $r$ denotes reaction runs with deuteron beam.  These label the yields, the beam energies, and the integrated beam flux on target.  Subscripts indicate reactions where needed.  The yields of interest are as follows:
\begin{eqnarray}
N_{\sdpp}^p\,=& n^pt_{d1}\Delta\Omega_A\sigma_{\sdpp}(E^p,\theta_{A}) \\
N_{\saupp}^p\,=& n^pt_{g1}\Delta\Omega_C\sigma_{\saupp}(E^p,\theta_{C}) \\
N_{\ddd}^d\,=& n^dt_{d1}\Delta\Omega_B\sigma_{\ddd}(E^d,\theta_{B}) \\
N_{\saudd}^d\,=& n^dt_{g1}\Delta\Omega_C\sigma_{\saudd}(E^d,\theta_{C}) \\
N_{\sddp}^r\,=& n^rt_{d2}\Delta\Omega_A\sigma_{\sddp}(E^r,\theta_{A}) \\
N_{\ddd}^r\,=& n^rt_{g2}\Delta\Omega_B\sigma_{\ddd}(E^r,\theta_{B})
\end{eqnarray}

The solution for the $\sddp$ differential cross section is given by the following:
\begin{equation}
\centering
\begin{array}{lcl}
\multicolumn{2}{l}{\displaystyle \lefteqn{\displaystyle \sigma_{\sddp}(E^r,\theta_A)=\sigma_{\dpp}(E^p,\theta)\times}}&{}\\\\
&&\displaystyle\frac{N_{\sddp}^r}{N_{\ddd}^r}
   \frac{N_{\ddd}^d}{N_{\saudd}^d}\frac{N_{\saupp}^p}{N_{\sdpp}^p}
   \frac{\sigma_{\saudd}(E^d,\theta_{C})}{\sigma_{\saupp}(E^p,\theta_{C})}\,.
\end{array}
\label{eq:alphas}
\end{equation}
In order to analyze the data one run at a time, we break this solution into the normalization factors $\alpha$ corresponding to the three different run types:
\begin{eqnarray}
\alpha^p&=& \frac{N_{\sdpp}^p}{N_{\saupp}^p}
\frac{\sigma_{\saupp}(E^p,\theta_{C})}{\sigma_{\dpp}(E^p,\theta_A)}
\label{eq:alphap}\\
\alpha^d&=& \frac{N_{\ddd}^d}{N_{\saudd}^d}\sigma_{\saudd}(E^d,\theta_C)
\label{eq:alphad}\\
\alpha^r&=& \frac{N_{\sddp}^r}{N_{\ddd}^r}
\label{eq:alphar}
\end{eqnarray}
Now we have simply
\begin{equation}
\sigma_{\sddp}(E^r,\theta_A)=\frac{\alpha^r\alpha^d}{\alpha^p}\,.
\label{eq:dsigalpha}
\end{equation}
By substituting the yields with the systematic parameters on which they depend, these factors can also be written in the following way:
\begin{eqnarray}
\alpha^p&=&\frac{\Delta\Omega_A}{\Delta\Omega_C}\frac{t_{d1}}{t_{g1}}
\label{eq:physalphap}\\
\alpha^d&=&\frac{\Delta\Omega_B}{\Delta\Omega_C}\frac{t_{d1}}{t_{g1}}\sigma_{\ddd}(E^r,\theta_B)
\label{eq:physalphad}\\
\alpha^r&=&\frac{\Delta\Omega_A}{\Delta\Omega_B}\frac{\sigma_{\sddp}(E^r,\theta_A)}
  {\sigma_{\ddd}(E^r,\theta_B)}
\label{eq:physalphar},    
\end{eqnarray}
which makes evident the solid angle and target thickness cancellations in Eq.~\ref{eq:dsigalpha}\@. 

To calculate \alp, differential cross-section values were needed for $p$--$d$ elastic scattering.  Theoretical calculations were provided by Kievsky {\it et al.}~\cite{kie95}~\cite{kievpdprivate} for several selected energies in the range of interest for the present work.   Two-dimensional polynomial interpolations were used to obtain values for the precise angles and energies required.  The error associated with the interpolation is negligible.

All cross sections for scattering on gold used in our analysis were calculated from the Rutherford scattering formula with electron-screening corrections obtained from Ref.~\cite{huttel83}\@.  At the lower energies, the screening corrections were about 1\% of the the values.

\begin{figure}[htb!] 
\centering
\includegraphics[width=\figwidth]{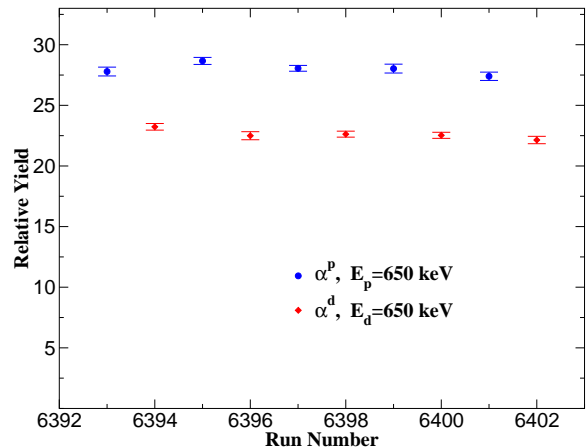}
\caption{(Color) Relative yields measured in beam-switching normalization runs: $\alpha^p$ and $\alpha^d$ are the normalized \sdpp and \ddd yields as defined in Eqs.~\ref{eq:alphap} and~\ref{eq:alphad}\@.}
\label{fig:anal:pdnorm}
\end{figure}

The error due to target thickness variation can be seen by plotting the values of $\alpha^p$ and $\alpha^d$ as a function of run number.  Run numbers were incremented consecutively as beams were switched between proton and deuteron beams. 
The results for the ~650~keV normalization are shown in Fig.~\ref{fig:anal:pdnorm}\@.  The two factors, having different physical meanings, are shown in relative units.  The relative stabilities of the two measurements is apparent.

\subsection{Angular Distributions}
\label{sec:analysis:dsdo}

Deadtime-corrected yields for all rotating detectors were normalized by dividing them by an appropriate combination monitor-detector yields.  The resulting normalized data was sensitive only to the differential cross sections of the reactions and an overall normalization factor.
The relative solid angle normalizations of the rotating detectors could be determined from cross-calibration data taken for that purpose, but ultimately these normalizations were left as free parameters as discussed in Sec.~\ref{sec:globalfit}\@.
With four monitor detectors there was sufficient redundancy to use an automated algorithm to reject monitors which were blocked or partially blocked in a particular chamber geometry.   Angle and solid-angle differences between the four monitors required the relative normalizations of the monitor yields to be determined for every energy.  The full data set of all reaction runs at each energy was used for this determination.  Iterations were made alternating between the relative normalization procedure and the monitor rejection algorithm.  

\begin{figure}[htb!]\centering
\includegraphics[width=\figwidth]{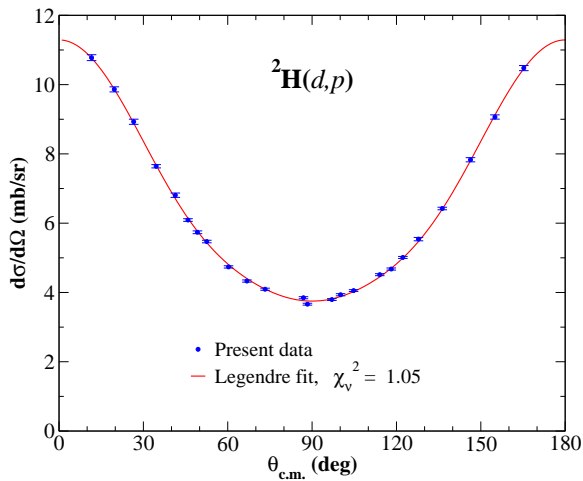}
\caption{(Color online) Normalized angular distribution of \ddp differential cross section at $E_d=650$~keV.  The data points shown are the weighted means of all data runs at the same angle.  The curve is a fit of even order Legendre polynomials up to $P_4$.}
    \label{fig:ddpmean660}
\end{figure}

Fig.~\ref{fig:ddpmean660} shows the resulting angular distribution for the \ddp reaction at $E_d=650$~keV with a fit to even-order Legendre polynomials in center-of-mass angle. 

\subsection{Low Energy Normalizations}
\label{sec:analysis:crunch:lownorm}\
Once normalized angular distributions were derived at $E_d=480$~keV, the energy cross-normalization data could be used to normalize the low energy angular distributions at fixed angles.   Again, for simplicity, we will consider just one reaction detector R.  Only the \ddp and \ddn reaction peaks were useable for the cross-normalizations.  
The reaction yields in R were normalized only to the elastic-scattering yields on gold in the Rutherford-scattering monitor, again labeled detector C, and to the appropriate Rutherford cross section.  The normalized yield \ax is defined as 
\begin{equation}
\ax=\frac{N_{\ddp}^d}{N_{\saudd}^d}\sigma_{\saudd}(E^d,\theta_C)\,.
\label{eq:alphax}
\end{equation}
This ratio can also be written as: 
\begin{equation}
\ax=\frac{\Delta\Omega_R}{\Delta\Omega_C}\frac{t_{d}}{t_{g}}\sigma_{\ddp}(E^d,\theta_B)\,,
\label{eq:physalphax}
\end{equation}
using the same notation as in previous sections.

If this ratio is determined for data taken at two deuteron energies, 480~keV, and for example, 120~keV, using the same target, then we find \\
\begin{equation}
\frac{\ax(E_d=120~{\rm keV})}{\ax(E_d=480~{\rm keV})}=\frac{\sigma_{ddp}( E_d=120{\rm ~keV},\theta)}
     {\sigma_{ddp}(E_d=480~{\rm keV},\theta)}\,.
\end{equation}

Low-energy \ddhet reaction cross sections can then be determined by relating the \ddhet and \ddp yields in the plentiful angular distribution data sets.

\subsection{Integrated Cross-Section Global Analysis}
\label{sec:enchilada}
\label{sec:globalfit}

The framework described in Sec.~\ref{chap:analysis} is sufficient for determining total-cross section results for both reactions at all observed energies. However, in the previous sections the situation was oversimplified by considering that each data type was observed using one detector and considering only one reaction at a time.  In this algebraic framework and without these simplifications, it is very difficult to fully utilize all the data from all detectors in the normalization process while accounting for the correlations involved.

To list one example, $p$--$d$ elastic-scattering  data and the normalization reaction data exist for two detector sets at two different angles.  In order for solid angles to vanish in the formalism of Sec.~\ref{chap:analysis}, the same detector set must be used in the analysis of the reaction data as for the proton elastic-scattering data.  Thus data from each detector set must be treated as an independent measurement.  This actually produces complications.  The fit to the angular distribution data does not directly produce discrete values as a function of angle; it produces Legendre coefficients and errors in those Legendre coefficients.   The values derived for the differential cross sections at particular angles have uncertainties which are highly correlated through the Legendre parameterization.  Such correlations in the analysis inputs ultimately complicate the calculation of the uncertainty of a single result.

We were also interested in calculating multiple related results, cross-sections for different reactions and different energies.  The results themselves have uncertainties which are correlated with each other.  This complicates the use of these values in future calculations such as the deuterium abundance prediction of a BBN network calculation.   In the present case, strong correlations in energy exist because the low energy data were normalized directly to high energy data.  Strong correlations between cross-section uncertainties for the two different reactions exist because these different reaction yields were normalized to the same data.  The uncertainty correlations in the cross sections must be quantified.

We have addressed these correlation issues by avoiding an algebraic analysis as much as reasonably possible.  We performed a $\chi^2$ fit of the energy-dependent differential cross sections.  These were fit to all the normalization and angular distribution data for both reactions at all energies, all in a single and simultaneous fit.  This procedure determined uncertainties for all parameters using all available data which may directly or indirectly affect their values. It also allowed the correlations in the uncertainties in those parameters to be measured using the usual parameter variation techniques of $\chi^2$ fitting.

To perform this fit we produced a realistic parameterization to describe all segments of the experiment.  The normalized data values which were input into the fit were the $\rm \alpha\,$'s of Eqs.~\ref{eq:alphap}-\ref{eq:alphar} and~\ref{eq:alphax}\@. The fit parameters represented the unknown values in the physical representations given in Eqs.~\ref{eq:physalphap}-\ref{eq:physalphar} and~\ref{eq:physalphax}\@.
The fit thus included a parameterization of the energy dependent differential cross sections along with all needed systematic parameters specific to the given data types.  Actually, many of these variables which cancelled out of the algebraic analysis, such as target thicknesses, were individually undetermined.  In most cases the appropriate fit parameters were carefully chosen to represent well determined products and ratios of these variables.  
We will define here only the parameters of interest.  The \ddn and \ddp differential cross sections can be described in the center-of-mass frame in terms of even Legendre polynomials $P_n(\theta)$ with coefficients $a_{n,ik}$ for the $n\,$'th Legendre polynomial, the $i\,$'th energy and the $k\,$'th reaction.  To improve the minimization behavior we defined the parameters $b_{n,ik}$ such that
\begin{equation}
b_{n,ik}=\frac{a_{n,ik}}{a_{0,ik}}\,.
\end{equation}
The differential cross sections are then parameterized as
\begin{equation}
\begin{array}{lcl}
\multicolumn{2}{l}{
\lefteqn{\sigma_k(E_i,\theta_{\rm c.m.})=}}\\&& a_{0,ik}P_0(\theta_{\rm c.m.})+\sum\limits_{m=1}^{M}b_{2m,ik}a_{0,ik}P_{2m}(\theta_{\rm c.m.})\,,
\label{eq:discretepar}
\end{array}
\end{equation}
where $2M$ is the highest order of Legendre polynomial used.  
The integrated cross sections are then given by 
\begin{equation}
\sigma_k(E_i)=4{\pi}a_{0,ik}\,.
\end{equation}

We also performed a fit to a parameterization which is continuous in energy.  This has many advantages, but also some complications.  We will not present the full results or description of that fit here; it is described in detail in Ref.~\cite{leonardthesis}\@.

\subsection{Error Analysis}
\label{sec:anal:error}

The uncertainties for the normalized data input into the fits were calculated by first-order propagation of the uncertainties in the peak sums.  Correlations arising from shared monitor normalization counts were not taken into account because the monitor yields generally contributed a relatively small amount to the overall statistical uncertainties.   The final uncertainties and correlations in the cross-section results
are determined directly from the error matrix.  We will explain this here in more detail and will discuss procedures applied to quantify data scatter and certain systematic errors.

\subsubsection{The Error Matrix}
\label{sec:anal:error:matrix}
The total cross section at each energy corresponds directly to one of the fit parameters.  The $\rm 1 \sigma$ uncertainty $\varsigma$ in the cross section is simply the uncertainty in the corresponding parameter multiplied by a factor of $4\pi$:
\begin{equation}
\varsigma(\sigma_k(E_i))=4{\pi}\varsigma({a_{0,ik}})\,,
\end{equation}
where $\varsigma(a_{0,ik})$ is given by the square root of the diagonal element of the covariant error matrix corresponding to the parameter $a_{0,ik}$.

\subsubsection{Quantifying Scatter}
\label{sec:error:scatter}

If for some fit, $\chi^2$ per degree of freedom, $\chi^2_{\nu}$, is not 1, then error matrix has little meaning.  Some amount of unaccounted-for scatter is expected in the data for various reasons,  
the largest coming from fluctuations in the target thicknesses during the beam-switching procedure.  In the end, $\chi_{\nu}^2$ for the entire fit is just over 2.

We implemented a procedure to address this issue by quantifying the scatter in the data and adding an appropriate amount of uncertainty to the data.  This was done by iteratively adding in quadrature enough uncertainty to points in each of the data sets in order to make $\chi^2$ per datum equal to 1 for that data set.  By adding a constant fractional error rather than multiplying the uncertainties, we were able to make better use of the full data set.  
This is a result of not over de-emphasizing points having uncertainties which were already large compared to the missing uncertainty.   For a single parameter measurement, in the case where the fraction of the uncertainty which was initially quantified approaches zero, this additive method reduces to the familiar technique of measuring the standard error in the mean.  

\subsubsection{Angle and Energy Uncertainties}
\label{sec:anal:error:zero} 

The effects of detector-angle and energy uncertainties were studied by perturbing the input angle and energy values of all data in various ways and then reproducing the entire analysis for each set of perturbations.  This is similar in principle to the uncertainty analysis performed by the fitting software, but for technical reasons these parameters could not easily be fully parameterized within our analysis framework.  

The detector angles were calibrated and set with an uncertainty of 0.1\degrees.
Fractional uncertainties arising from angle determinations were found to be on the scale of $2\times10^{-3}$ or less and were neglected.  Uncertainties arising from energy determinations are larger, as much as one or two percent at the lowest energies, and fall to a small fraction of a percent at the highest energies.  

In order to consolidate these uncertainties into the error matrix, the effect of the energy perturbations on the differential cross section parameters was itself parameterized and then fed back into the final fit.  The fit was then re-minimized.  Free parameters and constraint terms were included to allow the energies to vary somewhat in the fit.  This approach did not allow the fit to explicitly explore the effects of these perturbations on subtleties such as changes in center-of-mass angles of detectors.  We emphasize "explicitly" because such subtleties were in fact accounted for when originally parameterizing the effects of the perturbations. For this reason, all freedom in the final results arising from such effects was still quantified and reflected in the error matrix.  The increase in the uncertainties calculated by this fitting procedure was in excellent agreement with quadrature addition of the tabulated energy-related uncertainties to the original uncertainties in the fit.

\subsubsection{Peak-Fitting Uncertainties}
\label{sec:fitting_error}  

Some ambiguity existed in the fits of the Rutherford backscattering spectra for deuterons scattering from gold at the lowest energies.   It was generally unnecessary to account for correlations arising from uncertainties in monitor detector yields.   However, because of the systematic nature of this normalization peak and the non-negligible uncertainty in the fit, this rationale became invalid.  We added extra normalization parameters for these data in the global fit.  Corresponding constraint terms were added to the $\chi^2$ sum constraining these parameters to be near one.  The uncertainties assigned to the constraint terms were 3\% for the 120~keV normalization parameter and to 1\% for all other energies at or below 320~keV\@.

\section{Results, Implications, and Conclusions}
\label{chap:results}

Here we present our data and address the effect these data will have on Big-Bang Nucleosynthesis calculations.
The cross-section results obtained from the parameterization described in Eq.~\ref{eq:discretepar} are given in Table~\ref{tab:totalsfact}\@.   The coefficients of the Legendre expansions of the differential cross-sections are given in Table~\ref{tab:dsigs}\@.  The errors shown are 1$\sigma$ equivalents which include statistical and background uncertainties, energy uncertainties discussed in Sec.~\ref{sec:anal:error:zero}, fitting uncertainties described in Sec.~\ref{sec:fitting_error}, and which account for scatter in the data sets used to derive the total cross sections as explained in Sec.~\ref{sec:error:scatter}\@.  The uncertainties are generally around  the 2\% level except at the lowest energies where peak fitting and energy uncertainties become somewhat larger.   An overall scale error of 1\% is estimated arising from uncertainties in the $p$--$d$ elastic-scattering cross section to which these data were normalized.  This normalization error is not included in the uncertainties.  Finally, the error matrix for the total cross-section parameters is given in Table~\ref{tab:discmat}\@.  The complete error matrix for the Legendre coefficients is too large to present here and is not relevant to our primary goal of providing BBN inputs.

\begin{figure*}[htbp]\centering
\includegraphics[width=\singlewidth]{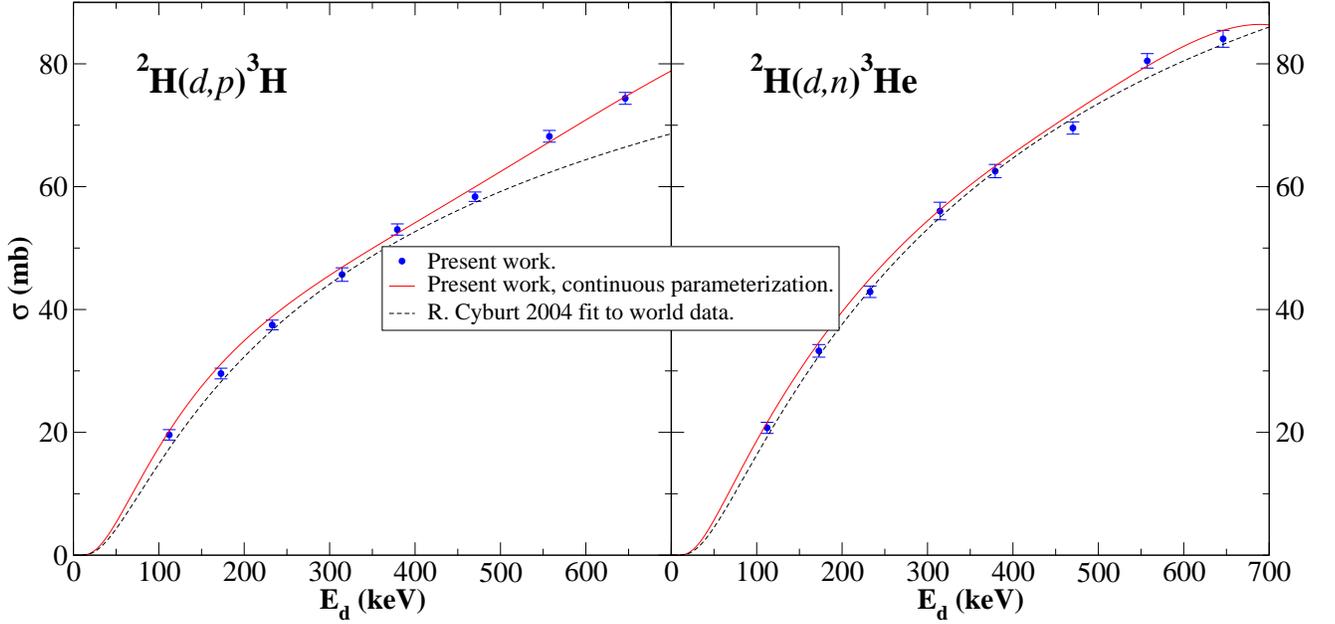}
\caption[Integrated cross-section results for \ddp and \ddn.]{(Color online) Integrated cross-section results for \ddp and \ddn\@.  The points are the results from the discrete parameterization of Eq.~\ref{eq:discretepar} and the curves are the results of the continuous parameterization.  The error bars are 1$\sigma$ uncertainties including statistical uncertainties as well as  fitting and energy uncertainties.  Details are discussed in Sec.~\ref{sec:anal:error}\@.}
\label{fig:bothsigs}
\end{figure*}

The cross sections of Table~\ref{tab:totalsfact} are plotted in Fig.~\ref{fig:bothsigs}, along with the continuous parameterization mentioned in Sec.~\ref{sec:enchilada} and with the recent cross section compilation of Cyburt~\cite{cyburtprd04}\@.  Some scatter within the uncertainties is detectable, which is expected since our statistical and systematic errors are of comparable magnitudes.    The continuous parameterization appears to be systematically somewhat higher than the results of the discrete fit.  This is not surprising since cross sections for the lowest four energies are all normalized directly to the differential cross sections at 480~keV\@. These data thus inherit all of the uncertainty of the 480~keV points, but in a systematic manner.  The continuous fit is then free to systematically renormalize these points within constraints.  This systematic uncertainty is quantified in the error matrix of the discrete fit.

The Cyburt compilation agrees well with the present data except at the high-energy end  the \ddp reaction cross sections.  At the lowest energy we can compare directly to the high-precision data of Brown and Jarmie~\cite{browndd}, which falls 11\% and 8\% below the present work for \ddp and \ddn respectively, corresponding to discrepancies of 2.1 $\sigma$ and  1.6 $\sigma$ respectively when systematic and statistical uncertainties of both experiments are included.  The data of Greife {\it et al.}~\cite{greife}, having uncertainties of only about 2.8\%, agree well with our results at higher energies, up to 256~keV, for both reactions.  

There have been several published compilations and analyses of BBN data and network calculations and we cannot review or even acknowledge all of the major efforts here.    Most, including Refs.~\cite{cuoco,cyburt,redux,Des05}, have focused a significant amount of attention on consistency of cosmological observables including primordial abundances and CMBR observations.  Many, including some which have now acquired benchmark status, have used Monte Carlo techniques to analyze the relationship between the uncertainties in the reaction data to the uncertainties in the astrophysical constraints~\cite{skm,romanelli,nollettburles}\@.  The work of the latter produced estimates of energy regions having the highest sensitivities to cross-section data, estimates which inspired the present work.  Finally there are even some works that have used traditional first-order error propagation to understand the effects of the data on network predictions.  For example, Fiorentini {\it et al.}~\cite{fior} achieved impressively good agreement with the computationally expensive techniques used elsewhere and have the advantage of producing simple functional understandings of various uncertainty relationships via tabulated derivatives.  

The NACRE compilation~\cite{NACRE} has become widely accepted as a standard in adopted values for reaction-rates of nuclear reactions of astrophysical significance.  NACRE does not address network calculations since it is intended as a broad resource for general astrophysical use.  It has though become the data source of choice for several BBN analyses~\cite{cuoco,flam,Coc04}\@.

The NACRE reaction-rate values cannot be compared directly to the cross-section data of the present work.  The reaction rate is an integral of the total cross section weighted by a Maxwellian distribution and taken over all energies.  In order to calculate the reaction-rate integral for the purpose of this comparison, we used the cross section results from the continuous parameterization of the present work.
For energies given in MeV and cross sections in barns, the reaction rate \rate in $\rm cm^3mol^{-1}s^{-1}$ is given by
\begin{equation}
\rate=3.731\times10^{10}\mu^{-1/2}T_9^{-3/2}\int_0^\infty\sigma E\,e^{-11.604\frac{E}{T_9}}dE\,,
\label{eq:rate}
\end{equation}
where $\mu$ is the reduced mass in amu and $T_9$ is the temperature in units of $\rm 10^9 K$.

For the integral to converge at temperatures relevant to BBN, $T_9<2$, it can be cut off at $E_{\rm c.m.}$ of about 2.5 MeV but not much lower.  Thus to calculate reaction rates accurately, we must include data at energies higher than those of the present work, although data at these energies contribute relatively little to the integral at the relevant temperatures.  For this purpose and to make a fair judgment of how the new data compare to the NACRE compilation, we have used the same data at high energies as those used in NACRE compilation, the data of Schulte {\it et al.}~\cite{schulte}\@.  We have included this data up to $ E_{\rm c.m.}=2.75~{\rm MeV}$.

\begin{figure}[htbp]\centering
\includegraphics[width=\figwidth]{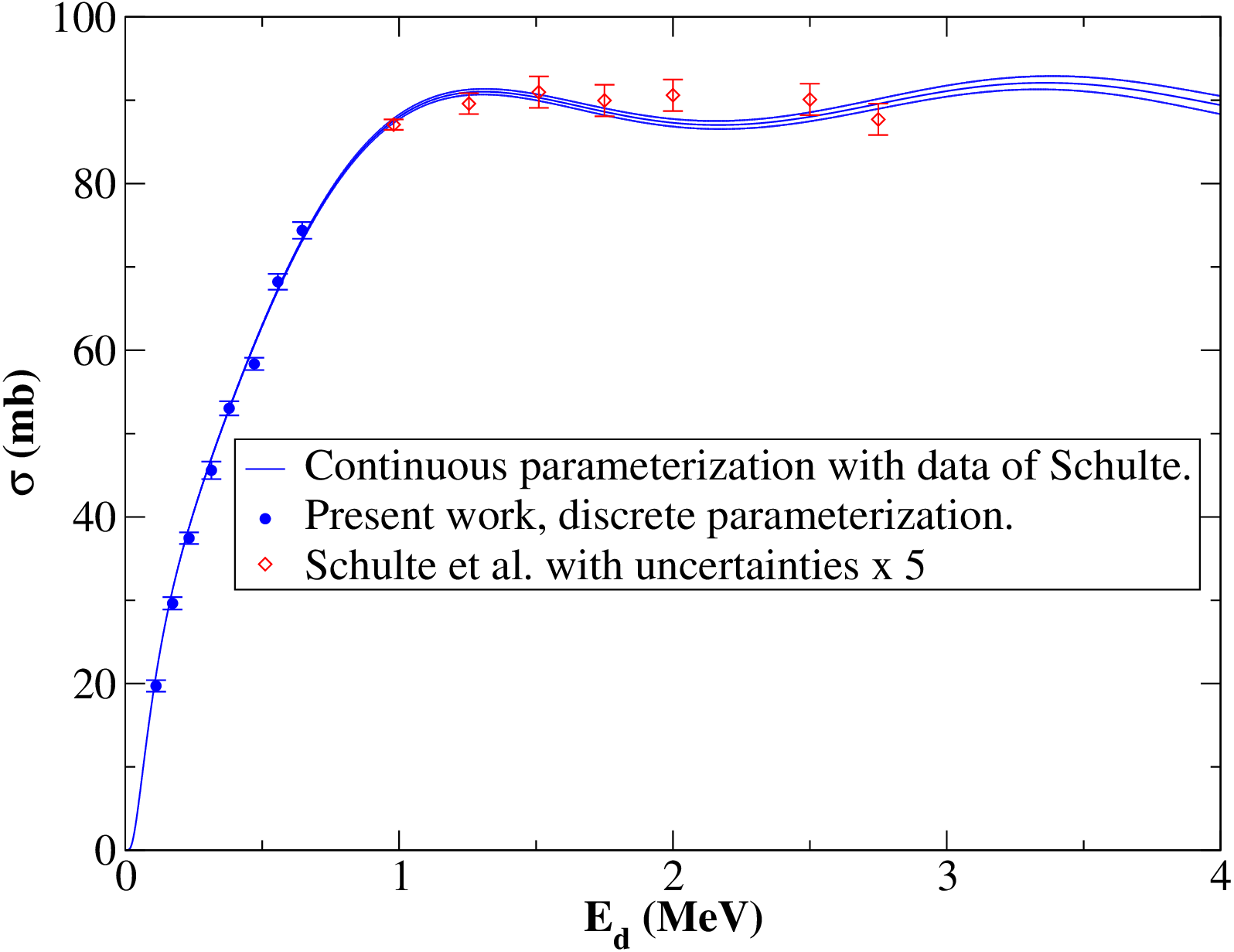}
\caption[Integrated cross-section result for \ddp of fit of continuous parameterization including integrated cross-section data of  {\it et al.}]{(Color) Integrated cross-section result for \ddp for fit of continuous parameterization including integrated cross-section data of Schulte {\it et al.}~\cite{schulte} The statistical uncertainties of Ref.~\cite{schulte} were multiplied by 5 in the figure and the fit.}
\label{fig:schultfit}
\end{figure}

\begin{figure}[htbp!]\centering
\includegraphics[width=\figwidth]{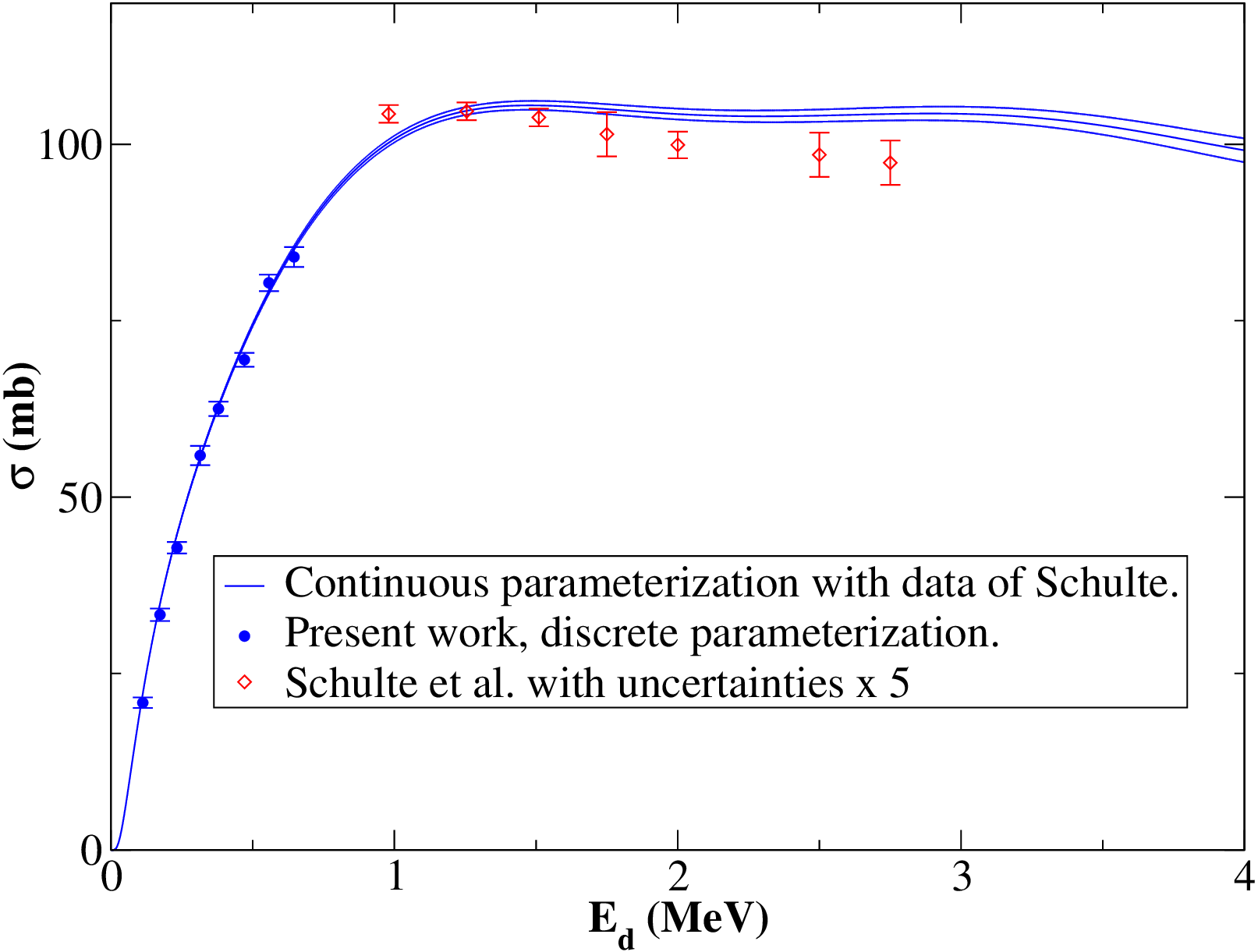}
\caption[Integrated cross-section result for \ddn of fit of continuous parameterization including integrated cross-section data of Schulte.]{(Color) Integrated cross-section result for \ddn for fit of continuous parameterization including integrated cross-section data of Schulte {\it et al.}~\cite{schulte} The statistical uncertainties of Ref.~\cite{schulte} were multiplied by 5 in the figure and the fit.}
\label{fig:heschultfit}
\end{figure}

These data were included in the continuous parameterization by using the total cross sections to appropriately constrain the zero order Legendre coefficients at the high energies.   The data of Ref.~\cite{schulte} have scatter which is significantly larger than their quoted statistical errors.  To prevent this data set from unduly constraining the fit, we have multiplied their statistical uncertainties by 5, leaving the smallest uncertainties in these data still below one percent of the value.  This multiplication is a coarse procedure but suffices for the present purpose.  
The results of these fits are shown in Figs.~\ref{fig:schultfit} and ~\ref{fig:heschultfit}\@.  The \ddn reaction curve does not follow the Schulte data impressively well.  This is probably due to a limitation of the parameterization in representing changes in curvature over these large energy regions.  It is not a significant problem for the purpose of constraining the tail of the reaction-rate integral.
 
The NACRE compilation gives coefficients for a quadratic cross-section function which was used in calculating the rates for the \ddn and \ddp reactions at ''low energies''\@.  NACRE does not specify exactly what is meant by low energies.  We have used these cross-section functions for energies below the present work in order to complete the integrals. As was the case for the high energy data, this low energy data contributes only a small amount to the integral for temperatures significant to BBN.
  
\begin{figure}[htbp!]\centering
\includegraphics[width=\figwidth]{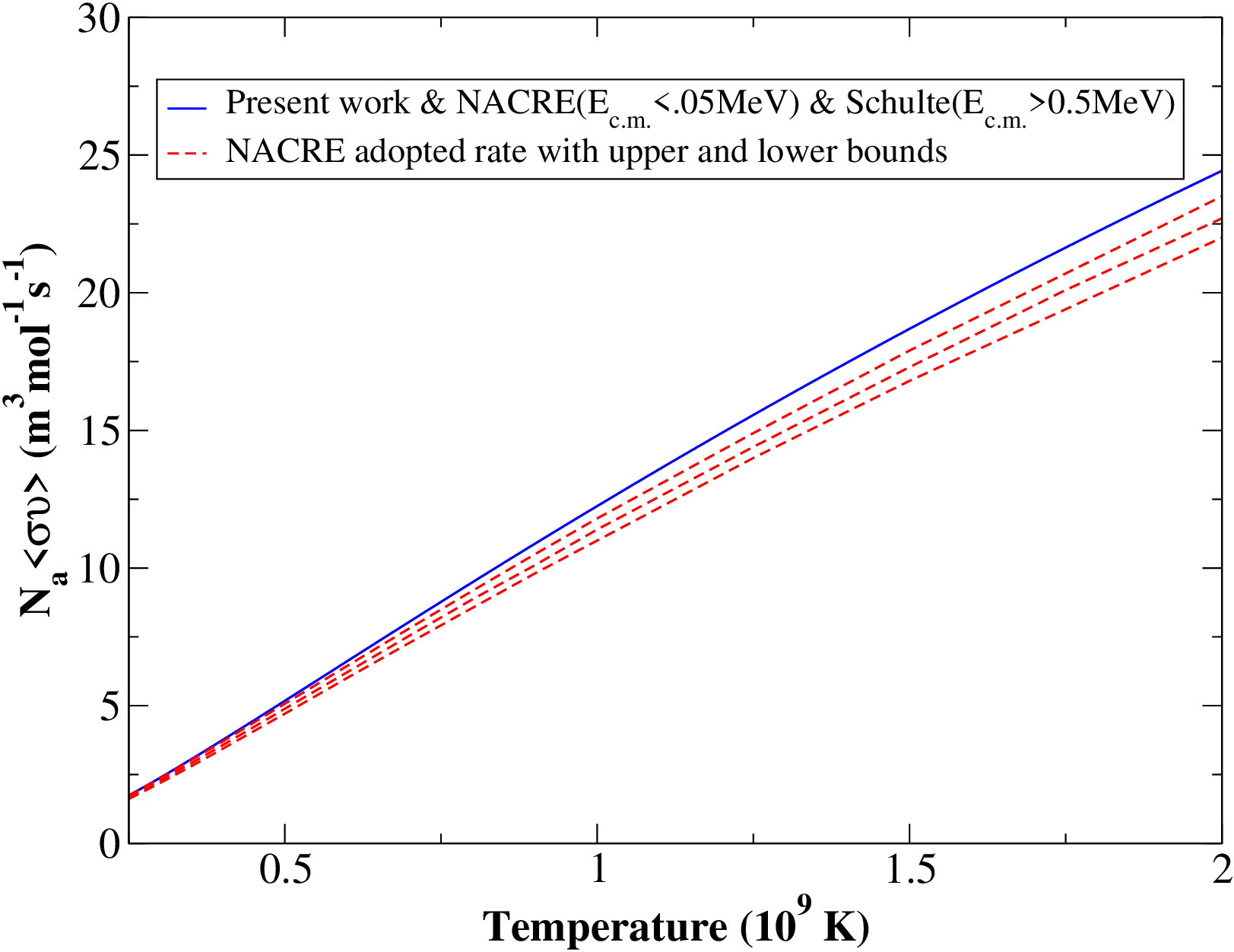}
\caption[Comparison of \ddp reaction rates of present work to NACRE compilation.]
{(Color online) Comparison of \ddp reaction rates of present work with the NACRE rate compilation. The results of the present work use the low energy cross-section fit of NACRE in the rate integral for $\rm {\it E}_{\rm c.m.}<0.05~MeV$.  Data of Ref.~\cite{schulte} are used to constrain the cross-section fit of the present work at energies above ${\it E}_{\rm c.m.}=0.5~{\rm MeV}$. }
\label{fig:ratecomp}
\end{figure}

\begin{figure}[htbp!]\centering
\includegraphics[width=\figwidth]{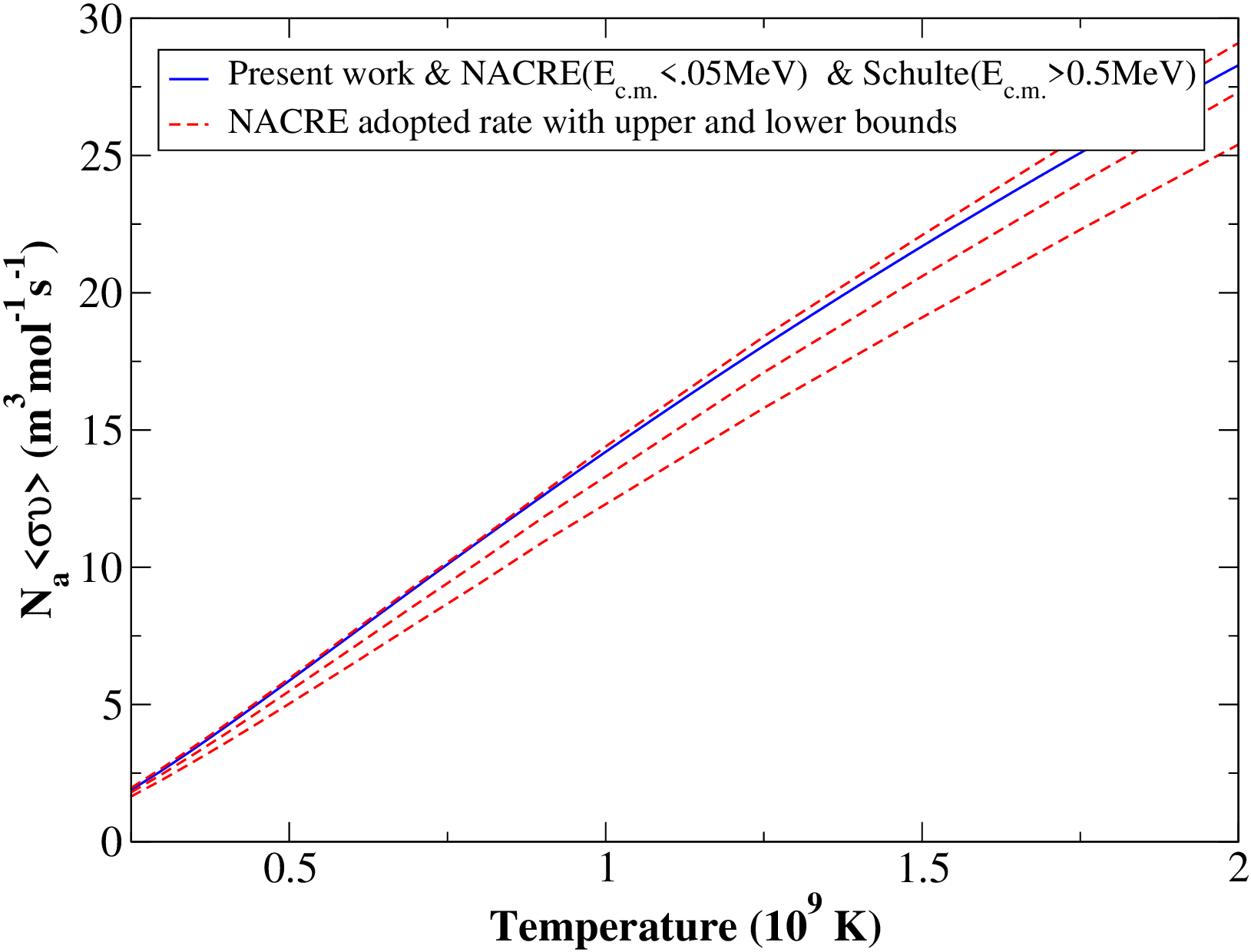}
\caption[Comparison of \ddn reaction rates of present work to NACRE compilation.]
{(Color online) Comparison of \ddn reaction rates of present work with NACRE rate compilation. The results of the present work use the low energy cross-section fit of NACRE in the rate integral for ${\it E}_{\rm c.m.}<0.05~MeV$.  Data of Ref.~\cite{schulte} are used to constrain the cross-section fit of the present work at energies above ${\it E}_{\rm c.m.}=0.5~{\rm MeV}$.}
\label{fig:heratecomp}
\end{figure}

Using these total cross-section curves, we then calculated the reaction rates by computing the right hand side of Eq.~\ref{eq:rate} with a high-energy cutoff on the integral at $ E_{\rm c.m.}=2.5{\rm MeV}$.  The results are compared to the NACRE reaction rates in Figs.~\ref{fig:ratecomp} and~\ref{fig:heratecomp}\@.   Judging by plots of abundances as a function of temperature for network calculations of Nollett and Burles~\cite{nollettburles}, all significant standard BBN occurs at temperatures near $T_9=1$.  At this temperature the rates derived using the present work are about 7\% higher than the NACRE rates for  both \ddn and \ddp\@.  These discrepancies are systematic, remaining at similar levels for a large range of temperatures.  Our derived \ddp rate is well beyond the NACRE quoted upper bounds, and for \ddn our value is marginally within the NACRE bounds.

It is important to emphasize that there is nothing inherently controversial about disagreement between the present work and NACRE or any other compilation.  The NACRE curves are derived from a limited set of data and in fact include no data between $E_d=325$~keV and $ E_d=2.0~{\rm MeV}$.  Furthermore, the data of Krauss {\it et al.}\@.~\cite{krauss} were the only data used in the  NACRE compilation in the energy range of 250 to 325 keV, and although these are respectable data, they have total uncertainties of about 8\% (see  Fig.~\ref{fig:ddndata}).   The NACRE compilation incorporates all data in these plots with the exception of the Ganeev data.   Our good agreement with Cyburt~\cite{cyburtprd04} is likely due to his inclusion of the Ganeev data set which, although it has large uncertainties, also agrees well with our data and covers the energy range with the least data available.  There was indeed a previous shortage of data and our results will greatly contribute to the cross-section information available in this energy region.

Many recent compilations such as that of Descouvemont {\it et al.}~\cite{Des04} used in the analysis of Coc {\it et al.}~\cite{Coc04}, as well as the compilation described by Serpico {\it et al.}~\cite{Ser04}, have a focus on BBN applications and strive to improve on the extrapolation techniques of the NACRE compilation in the BBN energy regions.  However, these compilations suffer from a lack of data in the same energy region as NACRE.    In spite of this, Serpico {\it et al.}  quote unusually low uncertainties at the levels of 1.3\% and 1\% for the \ddn and \ddp reaction rates respectively.  They do however acknowledge the severe lack of data and "strongly recommend a new experimental campaign" which we have now furthered.

Using the logarithmic derivatives tabulated in Ref.~\cite{fior}, we can estimate the differences in primordial abundances calculated using our new data vs. using the NACRE compilation.   These derivatives of abundances with respect to reaction rates are given in tables of coefficients of polynomial expansions in the baryon to photon ratio $\eta$.  For our estimates we use a value of $\eta$ of $6.14 \times 10^{10}$ from Ref.~\cite{cyburt} which was derived directly from an analysis of recent WMAP data performed in Ref.~\cite{spergelwmap}\@.  The values of the logarithmic derivatives calculated for $\eta=6.14 \times 10^{10}$ are given in Table~\ref{tab:logder}\@.
 
It is immediately evident that a roughly 7\% change in the cross sections will have essentially no effect on the primordial abundance of \hethree\@.   The relative signs of the derivatives for the two reactions have important significance.  Although the derivatives shown for the $\rm ^3He$ abundance are of significant magnitude, the changes in the abundance induced by the changes in the two reactions nearly cancel and the net effect is small.   However, because the two $D/H$ derivatives have the same sign as do the reaction rate changes inferred from the present work, 
the total change estimated in the deuterium abundance is roughly $-7$\%\@.  The change calculated for $\rm ^7Li/{\it H}$ is +5\% and is almost entirely attributable to the changes in the \ddn reaction rates.  

The lithium abundance observations are currently in severe disagreement with the Standard BBN results~\cite{kirkman}\@.  This 5\% increase in the predicted lithium abundance makes an unresolved problem slightly worse.  Currently the $\rm ^7Li$ abundances predicted using CMBR values of $\eta$ are at least a factor of 2 higher than the observed abundances~\cite{cuoco}\@.  The 5\% change is small compared with the uncertainties plaguing this comparison but it certainly does not improve the hopes of finding agreement between standard BBN and observed $\rm ^7Li$ abundances.  We look at this not as a failure but as an opportunity to explore possibilities of non-standard Big Bang Models and thus new physics.

The current observational value of the primordial deuterium abundance, $ {D/H}=2.78_{-0.38}^{+0.44}\times 10^{-5}$~\cite{kirkman}, obtained from QSO absorption spectra, is in good agreement with predictions of BBN using $\eta$ from the CMBR data, ${D/H}=2.56_{-0.24}^{+0.35}\times 10^{-5}$~\cite{cuoco} or $ {D/H}=2.55_{-0.20}^{+0.21}\times 10^{-5}$~\cite{cyburtprd04}\@.  However, as more data have arrived, the statistical uncertainty estimates on the abundance observations have not held up; the uncertainty is now dominated by scatter in the data.  The change which we predict in the $D/H$ value derived from BBN+CMBR will strain the current agreement slightly, but the results should still be well within the current mutual uncertainties.  The  reduced uncertainties of the \ddn and \ddp cross-sections from the present work, future improvements in precision of the \dpg cross-section measurements, and more QSO observations may soon provide significantly more stringent tests of this comparison, again yielding insight into the validity of the details of the standard BBN model.    

Although other aspects of the BBN picture are still limiting the comparisons of theory and experiment to around the level of 10\% or worse, the new data of the present work will pave the way for future developments in precision cosmology.  As other measurements are improved, BBN predictions will become sensitive to many details of the model including possible inhomogeneities and neutrino properties.   The new data should put to rest concerns and claims that the $d$--$d$ reaction rates may be inaccurate, and should ultimately result in significant modifications to the reaction rates used for current astrophysical applications including BBN\@.  The present data verify the limited, not always trusted, and often overlooked data sets which previously existed in this region of energy~\cite{ganeev} and with roughly 2\% to 3\% uncertainties, this work represents a significant improvement in precision, which along with improvements in other cross-sections, will translate directly into reductions in uncertainties of BBN predictions.  These data will truly help to usher in the ''new era''~\cite{cyburt} of precision cosmology.

\begin{acknowledgments}
    We thank Y. Tagishi and Kurt Fletcher for providing deuterated parapolyphenol for our targets, and Alejandro Kievsky for calculating $p$--$d$ scattering cross sections on request.  We also thank Ken Nollet and Scott Burles for valuable discussions and inputs regarding the necessary improvements in the BBN cross-sections, and for providing us with the original data of Ref.~\cite{greife} which can be difficult to find.  Finally we would like to acknowledge the support of the technical staff at Triangle Universities Nuclear Lab, especially John Dunham for providing dual-beam source configurations. This work was supported by the U.S. Department of Energy Grant No. DE-FG02-97ER41041.
\end{acknowledgments}

\begin{table}[htbp]\centering
\caption[Integrated cross-section results for discrete fit.]{Integrated cross-section results for discrete fit.  The corresponding S-factor error matrix is given in Table~\protect\ref{tab:discmat}\@.  The 1$\sigma$ uncertainties include statistical uncertainties from all parts of the present work as well as fitting uncertainties discussed in Sec.~\protect\ref{sec:anal:error}\@.}
 \begin{tabular}{ccc}
 \hline
 $E_d {\rm (keV)}$ &  $\rm \sigma_{\ddp}$ (mb) & $\rm \sigma_{\ddn}$ (mb) \\
 \hline
646.1 &   74.55 $\pm$     1.07 &    84.34 $\pm$     1.59 \\ 
557.3 &   68.02 $\pm$     0.96 &    80.22 $\pm$     1.18 \\ 
470.2 &   58.47 $\pm$     0.80 &    69.55 $\pm$     1.04 \\ 
379.2 &   53.09 $\pm$     0.94 &    62.60 $\pm$     1.12 \\ 
314.6 &   45.77 $\pm$     1.06 &    56.01 $\pm$     1.33 \\ 
232.8 &   37.65 $\pm$     0.82 &    43.13 $\pm$     0.96 \\ 
172.8 &   29.55 $\pm$     0.88 &    33.03 $\pm$     1.01 \\ 
112.2 &   19.77 $\pm$     0.85 &    21.07 $\pm$     0.92 \\ 
 \hline
 \end{tabular}
\label{tab:totalsfact}
\end{table}

\begin{table*}[htbp]\centering
\caption[Legendre coefficients for discrete fit of \ddp and \ddn data.]{Legendre coefficients for discrete fit of \ddp and \ddn data.  The 1$\sigma$ uncertainties include statistical uncertainties from all parts of the present work as well as fitting and energy uncertainties discussed in Sec.~\protect\ref{sec:anal:error}\@.  The coefficients are normalized to the $P_0$ coefficients.}
 \begin{tabular}{c|cc|cc}
 \hline
&\multicolumn{2}{c|}{\ddp} & \multicolumn {2}{c}{\ddn} \\
 $E_{d} {\rm (keV)}$ & $b_{2}$ & $b_{4}$ & $b_{2}$ & $b_{4}$ \\ 
 \hline
646.1 & 0.786 $\pm$  0.004 &  0.192 $\pm$  0.005 & 0.865 $\pm$  0.032 &  0.011 $\pm$  0.030 \\
557.3 & 0.737 $\pm$  0.004 &  0.199 $\pm$  0.006 & 0.856 $\pm$  0.011 &  0.156 $\pm$  0.013 \\
470.2 & 0.665 $\pm$  0.007 &  0.075 $\pm$  0.010 & 0.798 $\pm$  0.018 &  0.048 $\pm$  0.025 \\
379.2 & 0.629 $\pm$  0.007 &  0.094 $\pm$  0.008 & 0.830 $\pm$  0.016 &  0.112 $\pm$  0.019 \\
314.6 & 0.536 $\pm$  0.005 &  0.079 $\pm$  0.010 & 0.721 $\pm$  0.008 &  0.029 $\pm$  0.016 \\
232.8 & 0.472 $\pm$  0.005 &  0.013 $\pm$  0.010 & 0.663 $\pm$  0.007 &  0.011 $\pm$  0.011 \\
172.8 & 0.463 $\pm$  0.006 &  0.046 $\pm$  0.007 & 0.697 $\pm$  0.017 &  0.059 $\pm$  0.022 \\
112.2 & 0.358 $\pm$  0.012 &  0.024 $\pm$  0.008 & 0.578 $\pm$  0.010 &  0.016 $\pm$  0.008 \\
 \hline
 \end{tabular}
\label{tab:dsigs}
\end{table*}


\begin{table*}[htbp]\centering
\caption[Error Matrix for discrete fit.]{Error matrix for discrete fit.  the elements shown are for the zero order Legendre coefficients, the $a_0\,\rm{'s}$.  The two reactions \ddp and \ddn are labeled by p and n respectively and by their lab energies in keV\@.}
\small
 \begin{tabular}{c|cccccccc}
 \hline
 & p {\it E}=646.1 & p {\it E}=557.3 & p {\it E}=470.2 & p {\it E}=379.2 & p {\it E}=314.6 & p {\it E}=232.8 & p {\it E}=172.8 & p {\it E}=112.2 \\ 
 \hline
p {\it E}=646.1 & 0.72455$\rm \times 10^{-2}$ & 0.14698$\rm \times 10^{-3}$ & 0.17429$\rm \times 10^{-3}$ & 0.23506$\rm \times 10^{-3}$ & 0.26413$\rm \times 10^{-3}$ & 0.31480$\rm \times 10^{-3}$ & 0.37331$\rm \times 10^{-3}$ & 0.39490$\rm \times 10^{-3}$ \\ 
p {\it E}=557.3 & 0.14698$\rm \times 10^{-3}$ & 0.58695$\rm \times 10^{-2}$ & 0.20697$\rm \times 10^{-3}$ & 0.26821$\rm \times 10^{-3}$ & 0.31688$\rm \times 10^{-3}$ & 0.37043$\rm \times 10^{-3}$ & 0.43710$\rm \times 10^{-3}$ & 0.47199$\rm \times 10^{-3}$ \\ 
p {\it E}=470.2 & 0.17429$\rm \times 10^{-3}$ & 0.20697$\rm \times 10^{-3}$ & 0.40429$\rm \times 10^{-2}$ & 0.36346$\rm \times 10^{-3}$ & 0.34075$\rm \times 10^{-2}$ & 0.29271$\rm \times 10^{-2}$ & 0.24840$\rm \times 10^{-2}$ & 0.19097$\rm \times 10^{-2}$ \\ 
p {\it E}=379.2 & 0.23506$\rm \times 10^{-3}$ & 0.26821$\rm \times 10^{-3}$ & 0.36346$\rm \times 10^{-3}$ & 0.56466$\rm \times 10^{-2}$ & 0.55288$\rm \times 10^{-3}$ & 0.64357$\rm \times 10^{-3}$ & 0.75711$\rm \times 10^{-3}$ & 0.82970$\rm \times 10^{-3}$ \\ 
p {\it E}=314.6 & 0.26413$\rm \times 10^{-3}$ & 0.31688$\rm \times 10^{-3}$ & 0.34075$\rm \times 10^{-2}$ & 0.55288$\rm \times 10^{-3}$ & 0.70794$\rm \times 10^{-2}$ & 0.27062$\rm \times 10^{-2}$ & 0.23987$\rm \times 10^{-2}$ & 0.19782$\rm \times 10^{-2}$ \\ 
p {\it E}=232.8 & 0.31480$\rm \times 10^{-3}$ & 0.37043$\rm \times 10^{-3}$ & 0.29271$\rm \times 10^{-2}$ & 0.64357$\rm \times 10^{-3}$ & 0.27062$\rm \times 10^{-2}$ & 0.42573$\rm \times 10^{-2}$ & 0.22736$\rm \times 10^{-2}$ & 0.19649$\rm \times 10^{-2}$ \\ 
p {\it E}=172.8 & 0.37331$\rm \times 10^{-3}$ & 0.43710$\rm \times 10^{-3}$ & 0.24840$\rm \times 10^{-2}$ & 0.75711$\rm \times 10^{-3}$ & 0.23987$\rm \times 10^{-2}$ & 0.22736$\rm \times 10^{-2}$ & 0.49179$\rm \times 10^{-2}$ & 0.19986$\rm \times 10^{-2}$ \\ 
p {\it E}=112.2 & 0.39490$\rm \times 10^{-3}$ & 0.47199$\rm \times 10^{-3}$ & 0.19097$\rm \times 10^{-2}$ & 0.82970$\rm \times 10^{-3}$ & 0.19782$\rm \times 10^{-2}$ & 0.19649$\rm \times 10^{-2}$ & 0.19986$\rm \times 10^{-2}$ & 0.46104$\rm \times 10^{-2}$ \\ 
n {\it E}=646.1 & 0.79321$\rm \times 10^{-2}$ & 0.17280$\rm \times 10^{-3}$ & 0.19870$\rm \times 10^{-3}$ & 0.26475$\rm \times 10^{-3}$ & 0.30188$\rm \times 10^{-3}$ & 0.35790$\rm \times 10^{-3}$ & 0.42390$\rm \times 10^{-3}$ & 0.45124$\rm \times 10^{-3}$ \\ 
n {\it E}=557.3 & 0.17579$\rm \times 10^{-3}$ & 0.68319$\rm \times 10^{-2}$ & 0.24401$\rm \times 10^{-3}$ & 0.31657$\rm \times 10^{-3}$ & 0.37347$\rm \times 10^{-3}$ & 0.43684$\rm \times 10^{-3}$ & 0.51555$\rm \times 10^{-3}$ & 0.55638$\rm \times 10^{-3}$ \\ 
n {\it E}=470.2 & 0.20782$\rm \times 10^{-3}$ & 0.24633$\rm \times 10^{-3}$ & 0.45101$\rm \times 10^{-2}$ & 0.43104$\rm \times 10^{-3}$ & 0.37825$\rm \times 10^{-2}$ & 0.32794$\rm \times 10^{-2}$ & 0.28069$\rm \times 10^{-2}$ & 0.21721$\rm \times 10^{-2}$ \\ 
n {\it E}=379.2 & 0.27698$\rm \times 10^{-3}$ & 0.31625$\rm \times 10^{-3}$ & 0.42712$\rm \times 10^{-3}$ & 0.62301$\rm \times 10^{-2}$ & 0.65042$\rm \times 10^{-3}$ & 0.75900$\rm \times 10^{-3}$ & 0.89207$\rm \times 10^{-3}$ & 0.97730$\rm \times 10^{-3}$ \\ 
n {\it E}=314.6 & 0.32327$\rm \times 10^{-3}$ & 0.38773$\rm \times 10^{-3}$ & 0.41751$\rm \times 10^{-2}$ & 0.68081$\rm \times 10^{-3}$ & 0.85966$\rm \times 10^{-2}$ & 0.33147$\rm \times 10^{-2}$ & 0.29375$\rm \times 10^{-2}$ & 0.24227$\rm \times 10^{-2}$ \\ 
n {\it E}=232.8 & 0.36063$\rm \times 10^{-3}$ & 0.42425$\rm \times 10^{-3}$ & 0.33590$\rm \times 10^{-2}$ & 0.74001$\rm \times 10^{-3}$ & 0.31022$\rm \times 10^{-2}$ & 0.48744$\rm \times 10^{-2}$ & 0.26070$\rm \times 10^{-2}$ & 0.22529$\rm \times 10^{-2}$ \\ 
n {\it E}=172.8 & 0.41759$\rm \times 10^{-3}$ & 0.48870$\rm \times 10^{-3}$ & 0.27770$\rm \times 10^{-2}$ & 0.84616$\rm \times 10^{-3}$ & 0.26819$\rm \times 10^{-2}$ & 0.25418$\rm \times 10^{-2}$ & 0.54984$\rm \times 10^{-2}$ & 0.22346$\rm \times 10^{-2}$ \\ 
n {\it E}=112.2 & 0.42060$\rm \times 10^{-3}$ & 0.50305$\rm \times 10^{-3}$ & 0.20296$\rm \times 10^{-2}$ & 0.88470$\rm \times 10^{-3}$ & 0.21051$\rm \times 10^{-2}$ & 0.20917$\rm \times 10^{-2}$ & 0.21272$\rm \times 10^{-2}$ & 0.49029$\rm \times 10^{-2}$ \\ 
 \hline
 & n {\it E}=646.1 & n {\it E}=557.3 & n {\it E}=470.2 & n {\it E}=379.2 & n {\it E}=314.6 & n {\it E}=232.8 & n {\it E}=172.8 & n {\it E}=112.2 \\ 
 \hline
p {\it E}=646.1 & 0.79321$\rm \times 10^{-2}$ & 0.17579$\rm \times 10^{-3}$ & 0.20782$\rm \times 10^{-3}$ & 0.27698$\rm \times 10^{-3}$ & 0.32327$\rm \times 10^{-3}$ & 0.36063$\rm \times 10^{-3}$ & 0.41759$\rm \times 10^{-3}$ & 0.42060$\rm \times 10^{-3}$ \\ 
p {\it E}=557.3 & 0.17280$\rm \times 10^{-3}$ & 0.68319$\rm \times 10^{-2}$ & 0.24633$\rm \times 10^{-3}$ & 0.31625$\rm \times 10^{-3}$ & 0.38773$\rm \times 10^{-3}$ & 0.42425$\rm \times 10^{-3}$ & 0.48870$\rm \times 10^{-3}$ & 0.50305$\rm \times 10^{-3}$ \\ 
p {\it E}=470.2 & 0.19870$\rm \times 10^{-3}$ & 0.24401$\rm \times 10^{-3}$ & 0.45101$\rm \times 10^{-2}$ & 0.42712$\rm \times 10^{-3}$ & 0.41751$\rm \times 10^{-2}$ & 0.33590$\rm \times 10^{-2}$ & 0.27770$\rm \times 10^{-2}$ & 0.20296$\rm \times 10^{-2}$ \\ 
p {\it E}=379.2 & 0.26475$\rm \times 10^{-3}$ & 0.31657$\rm \times 10^{-3}$ & 0.43104$\rm \times 10^{-3}$ & 0.62301$\rm \times 10^{-2}$ & 0.68081$\rm \times 10^{-3}$ & 0.74001$\rm \times 10^{-3}$ & 0.84616$\rm \times 10^{-3}$ & 0.88470$\rm \times 10^{-3}$ \\ 
p {\it E}=314.6 & 0.30188$\rm \times 10^{-3}$ & 0.37347$\rm \times 10^{-3}$ & 0.37825$\rm \times 10^{-2}$ & 0.65042$\rm \times 10^{-3}$ & 0.85966$\rm \times 10^{-2}$ & 0.31022$\rm \times 10^{-2}$ & 0.26819$\rm \times 10^{-2}$ & 0.21051$\rm \times 10^{-2}$ \\ 
p {\it E}=232.8 & 0.35790$\rm \times 10^{-3}$ & 0.43684$\rm \times 10^{-3}$ & 0.32794$\rm \times 10^{-2}$ & 0.75900$\rm \times 10^{-3}$ & 0.33147$\rm \times 10^{-2}$ & 0.48744$\rm \times 10^{-2}$ & 0.25418$\rm \times 10^{-2}$ & 0.20917$\rm \times 10^{-2}$ \\ 
p {\it E}=172.8 & 0.42390$\rm \times 10^{-3}$ & 0.51555$\rm \times 10^{-3}$ & 0.28069$\rm \times 10^{-2}$ & 0.89207$\rm \times 10^{-3}$ & 0.29375$\rm \times 10^{-2}$ & 0.26070$\rm \times 10^{-2}$ & 0.54984$\rm \times 10^{-2}$ & 0.21272$\rm \times 10^{-2}$ \\ 
p {\it E}=112.2 & 0.45124$\rm \times 10^{-3}$ & 0.55638$\rm \times 10^{-3}$ & 0.21721$\rm \times 10^{-2}$ & 0.97730$\rm \times 10^{-3}$ & 0.24227$\rm \times 10^{-2}$ & 0.22529$\rm \times 10^{-2}$ & 0.22346$\rm \times 10^{-2}$ & 0.49029$\rm \times 10^{-2}$ \\ 
n {\it E}=646.1 & 0.16053$\rm \times 10^{-1}$ & 0.21150$\rm \times 10^{-3}$ & 0.23679$\rm \times 10^{-3}$ & 0.31202$\rm \times 10^{-3}$ & 0.36945$\rm \times 10^{-3}$ & 0.40998$\rm \times 10^{-3}$ & 0.47410$\rm \times 10^{-3}$ & 0.48070$\rm \times 10^{-3}$ \\ 
n {\it E}=557.3 & 0.21150$\rm \times 10^{-3}$ & 0.87689$\rm \times 10^{-2}$ & 0.29043$\rm \times 10^{-3}$ & 0.37325$\rm \times 10^{-3}$ & 0.45698$\rm \times 10^{-3}$ & 0.50031$\rm \times 10^{-3}$ & 0.57641$\rm \times 10^{-3}$ & 0.59298$\rm \times 10^{-3}$ \\ 
n {\it E}=470.2 & 0.23679$\rm \times 10^{-3}$ & 0.29043$\rm \times 10^{-3}$ & 0.68830$\rm \times 10^{-2}$ & 0.50724$\rm \times 10^{-3}$ & 0.46313$\rm \times 10^{-2}$ & 0.37600$\rm \times 10^{-2}$ & 0.31381$\rm \times 10^{-2}$ & 0.23088$\rm \times 10^{-2}$ \\ 
n {\it E}=379.2 & 0.31202$\rm \times 10^{-3}$ & 0.37325$\rm \times 10^{-3}$ & 0.50724$\rm \times 10^{-3}$ & 0.79473$\rm \times 10^{-2}$ & 0.80062$\rm \times 10^{-3}$ & 0.86980$\rm \times 10^{-3}$ & 0.99700$\rm \times 10^{-3}$ & 0.10421$\rm \times 10^{-2}$ \\ 
n {\it E}=314.6 & 0.36945$\rm \times 10^{-3}$ & 0.45698$\rm \times 10^{-3}$ & 0.46313$\rm \times 10^{-2}$ & 0.80062$\rm \times 10^{-3}$ & 0.11213$\rm \times 10^{-1}$ & 0.38031$\rm \times 10^{-2}$ & 0.32844$\rm \times 10^{-2}$ & 0.25781$\rm \times 10^{-2}$ \\ 
n {\it E}=232.8 & 0.40998$\rm \times 10^{-3}$ & 0.50031$\rm \times 10^{-3}$ & 0.37600$\rm \times 10^{-2}$ & 0.86980$\rm \times 10^{-3}$ & 0.38031$\rm \times 10^{-2}$ & 0.58079$\rm \times 10^{-2}$ & 0.29145$\rm \times 10^{-2}$ & 0.23981$\rm \times 10^{-2}$ \\ 
n {\it E}=172.8 & 0.47410$\rm \times 10^{-3}$ & 0.57641$\rm \times 10^{-3}$ & 0.31381$\rm \times 10^{-2}$ & 0.99700$\rm \times 10^{-3}$ & 0.32844$\rm \times 10^{-2}$ & 0.29145$\rm \times 10^{-2}$ & 0.64815$\rm \times 10^{-2}$ & 0.23776$\rm \times 10^{-2}$ \\ 
n {\it E}=112.2 & 0.48070$\rm \times 10^{-3}$ & 0.59298$\rm \times 10^{-3}$ & 0.23088$\rm \times 10^{-2}$ & 0.10421$\rm \times 10^{-2}$ & 0.25781$\rm \times 10^{-2}$ & 0.23981$\rm \times 10^{-2}$ & 0.23776$\rm \times 10^{-2}$ & 0.53203$\rm \times 10^{-2}$ \\ 
  \hline
 \end{tabular}
\normalsize
\label{tab:discmat}
\end{table*}

\begin{table}[htbp]\centering
\caption[Logarithmic derivatives of abundances with respect to reaction rates.]{
Logarithmic derivatives of abundances $Y$ with respect to reaction rates $R$ calculated from Ref.~\cite{fior}\@.}  
\begin{tabular}{ccc}
\hline
Abundance Ratio & $\frac{\partial ln Y_i}{\partial ln R_{\sddp}}$ 
&  $\frac{\partial ln Y_i}{\partial ln R_{\sddn}}$ \\
\hline
$D/H$  &  -0.46 & -0.53 \\
$\rm ^3He/{\it H}$   & -0.26 & 0.18 \\
$\rm ^4He$ mass fraction & 0.01 & 0.01 \\ 
$\rm ^7Li/{\it H}$ & 0.06 & 0.69 \\
\hline
\end{tabular}
\label{tab:logder}
\end{table}

\newpage
\bibliography{dd}
\end{document}